\documentclass[a4paper,fleqn]{cas-sc}

\usepackage[authoryear,longnamesfirst]{natbib}
\usepackage[utf8]{inputenc}
\usepackage{csvsimple}
\usepackage{subcaption}
\usepackage{threeparttable}

\usepackage{xurl}

\def\tsc#1{\csdef{#1}{\textsc{\lowercase{#1}}\xspace}}
\tsc{WGM}
\tsc{QE}

\begin{document}
\let\WriteBookmarks\relax
\def\floatpagepagefraction{1}
\def\textpagefraction{.001}

\shorttitle{GDP-GFCF Dynamics Across Global Economies}    

\shortauthors{A. Landowska, R. Kłopotek, D. Filip, K. Raczkowski}  

\title [mode = title]{GDP-GFCF Dynamics Across Global Economies: A Comparative Study of Panel Regressions and Random Forest}



%

\author[1]{Alina Landowska}[bioid=1, orcid=0000-0002-7966-8243]

\cormark[1]


\ead{alina.landowska@gmail.com}

\ead[url]{https://swps.pl/alina-landowska}

\credit{The conception and design of the study, analysis and interpretation of the results, conclusion, and policy implications} 

\affiliation[1]{organization={SWPS University},
            addressline={Chodakowska 19/31}, 
            city={Warsaw},
            postcode={03-815}, 
            country={Poland}}

\author[2]{Robert A. Kłopotek}[bioid=2, orcid=0000-0001-9783-4914]


\ead{r.klopotek@uksw.edu.pl}

\ead[url]{https://rklopotek.blog.uksw.edu.pl/o-mnie/}

\credit{Development and implementation of the Random Forest methodology} 

\affiliation[2]{organization={Cardinal Stefan Wyszyński University},
            addressline={Wóycickiego 1/3}, 
            city={Warsaw},
            postcode={01-938}, 
            country={Poland}}

\author[2]{Dariusz Filip}[bioid=3, orcid=0000-0002-6905-1004]


\ead{d.filip@uksw.edu.pl}

\ead[url]{https://wse.uksw.edu.pl/dariusz-filip/}

\credit{Data acquisition and methodology development for the panel regression analysis}

\author[2]{Konrad Raczkowski}[bioid=4, orcid=0000-0002-8546-2647]


\ead{k.raczkowski@uksw.edu.pl}

\ead[url]{https://wse.uksw.edu.pl/konrad-raczkowski/}

\credit{Critical revision and discussion of the study} 

\cortext[1]{Corresponding author}


\nonumnote{}

\begin{abstract}
This study examines the relationship between GDP growth and Gross Fixed Capital Formation (GFCF) across developed economies (G7, EU-15, OECD) and emerging markets (BRICS). We integrate Random Forest machine learning (non-linear regression) with traditional econometric models (linear regression) to better capture non-linear interactions in investment analysis. Our findings reveal that while GDP growth positively influences corporate investment, its impact varies significantly by region. Developed economies show stronger GDP-GFCF linkages due to stable financial systems, while emerging markets demonstrate weaker connections due to economic heterogeneity and structural constraints. Random Forest models indicate that GDP growth's importance is lower than suggested by traditional econometrics, with lagged GFCF emerging as the dominant predictor—confirming investment follows path-dependent patterns rather than short-term GDP fluctuations. Regional variations in investment drivers are substantial: taxation significantly influences developed economies but minimally affects BRICS, while unemployment strongly drives investment in BRICS but less so elsewhere. We introduce a parallelized p-value importance algorithm for Random Forest that enhances computational efficiency while maintaining statistical rigor through sequential testing methods (SPRT and SAPT). The research demonstrates that hybrid methodologies combining machine learning with econometric techniques provide more nuanced understanding of investment dynamics, supporting region-specific policy design and improving forecasting accuracy.
\end{abstract}


\begin{highlights}
\item GDP growth impacts GFCF differently: stronger in developed economies, weaker in BRICS
\item Beyond GDP: taxation matters in developed economies, unemployment drives BRICS investment
\item RF shows GDP less influential, lagged GFCF dominates in a path-dependent process
\item New statistical framework (SPRT \& SAPT) enhances machine learning variable importance testing
\item Hybrid models merge ML \& econometrics, tailoring policy to regional economic roles

\end{highlights}


\maketitle

\section{Introduction}\label{Intro}

Economic growth and private-sector investment are critical drivers of development, shaping employment, productivity, and long-term economic stability \citep{barro_economic_1989, jorgenson_capital_1963}. Among various indicators, Gross Fixed Capital Formation (GFCF) serves as a pivotal measure of corporate investment and its alignment with GDP growth \citep{de_long_equipment_1990}. Understanding the relationship between GDP growth and GFCF is essential for designing effective economic policies. However, traditional econometric methods often oversimplify this relationship, failing to capture its inherent non-linearities and the contextual dynamics of developed and emerging markets \citep{feldstein_domestic_1980}.

Existing studies predominantly rely on static or dynamic linear regression models to analyze GDP-GFCF dynamics, often overlooking the complexity of macroeconomic interactions and temporal dependencies \citep{blundell_initial_1998,feldstein_domestic_1980}. While linear approaches provide interpretable results, they may miss critical non-linear relationships and interactions among predictors \citep{breiman_random_2001}. Furthermore, emerging markets, such as BRICS, present unique challenges when it comes to conducting research — including data limitations, economic heterogeneity or non-robustness of the results obtained and alternative explanations — that remain underexplored in the literature \citep{xiang_time-varying_2021}.\\

This study seeks to bridge existing gaps by integrating traditional econometric models with advanced machine learning techniques to analyze the relationship between GDP-GFCF. Our goals are to: i. Investigates the influence of GDP growth on GFCF across developed and emerging markets \citep{devereux_taxes_1998} and other macroeconomic predictors, such as taxation, unemployment, and inflation, in shaping GFCF \citep{hall_tax_1967, pasara_causality_2020}; ii. Employs Random Forest (RF) to uncover non-linear interactions and rank predictor importance, providing a nuanced understanding of investment dynamics \cite{breiman_random_2001}; iii. Compare both methodologies, specifically linear regression and Random Forest. The hypotheses we test include:
\paragraph{[H1]}
GDP growth positively influences corporate investment (GFCF) across different regions, but its relative importance varies depending on the region and model specification.
\paragraph{[H2]}
The relationship between GDP growth and GFCF is stronger in emerging markets (BRICS) compared to developed markets (G7, OECD, EU-15), but other macroeconomic factors such as taxation and unemployment play equally or more significant roles.
\paragraph{[H3]}
Non-linear models, such as Random Forest, capture additional complexities in the relationship between GDP growth and GFCF that are not captured by linear static and dynamic regression models, highlighting the importance of other predictors like lagged GFCF.\\

A key contribution of this research is the novel use of RF models not for prediction but as a diagnostic tool to analyze and benchmark performance against traditional regression methods. This approach highlights non-linear relationships, identifies the dominant role of lagged GFCF, and addresses temporal dependencies that conventional models often fail to capture. By comparing RF with econometric models, the study illustrates the complementary strengths of these methodologies. Our contribution is a parallelized implementation of the p-value importance algorithm for Random Forests, leveraging multiprocessing to enhance computational efficiency while maintaining statistical rigor with methods like SPRT and SAPT, enabling scalable analysis of large macroeconomic datasets across regions. \\

The remainder of this paper is structured as follows: Section 2 reviews the relevant literature mainly on GDP growth and GFCF dynamics. Section 3 outlines the data and methodologies used. Section 4 presents the key results and analysis. Section 5 discusses the implications of the findings, while Section 6 addresses the study's limitations, and  Section 7 offers directions for future research. Finally, Section 8 concludes with a summary of the key contributions and insights.
\section{Background}\label{literature}
\subsection{Macroeconomic Factors influencing GFCF}

Higher GDP growth forecasts tend to stimulate greater GFCF, as businesses expand their capital investments to meet anticipated future demand. Conversely, GFCF itself is a key driver of long-term economic growth, playing an especially significant role in developing economies, where capital formation is critical for enhancing productivity and supporting sustainable development.


\cite{jorgenson_capital_1963} demonstrated that investment is significantly influenced by macroeconomic variables like GDP growth. According to \cite{le_foreign_2024}, an increase in gross capital formation (GCF), of which GFCF is a component, substantially boosts GDP growth; specifically, a 1\% rise in GCF leads to a 24.1\% increase in GDP. Similarly to \cite{pasara_causality_2020}, who identified a positive correlation between GCF and economic growth in South Africa.


Empirical evidence \citep{blanchard_lectures_1989, de_long_equipment_1990, world_bank_world_2018, serven_economic_1993} confirms the relationship between GDP growth and GFCF, especially in emerging markets. These studies demonstrate that higher GDP growth rates lead to significant increases in capital formation as firms expand to support growing demand. \cite{apergis_renewable_2010} further illustrate this relationship in the Commonwealth of Independent States (CIS), showing sector-specific responses to GDP growth. Similarly, \cite{xiang_time-varying_2021} explored time-varying impacts of GDP growth on investment in BRICS countries, reinforcing the correlation between economic expansion and capital formation, especially in capital-intensive sectors. Furthermore, in emerging markets, national savings significantly affect GFCF, as highlighted by the Feldstein-Horioka Puzzle \citep{feldstein_domestic_1980}. GDP growth stands out as one of the most significant predictors of corporate investment and GFCF, surpassing other macroeconomic variables such as the unemployment rate, tax rates, and the consumer price index.

GFCF, by investment in physical assets, can stimulate job creation by increasing the production capacity of firms and industries \citep{blanchard_regional_1992}. For instance, it has been demonstrated that as GDP and capital investments grow, labor markets benefit, leading to reduced unemployment rates. \cite{pasara_causality_2020} recently confirmed that there is a significant and positive relationship between unemployment and GCF, indicating that higher unemployment can lead to increased capital formation. On the flip side, high levels of unemployment can sometimes discourage capital investment. When unemployment rises, consumer demand often falls, making it less attractive for businesses to invest in new capital assets. Firms may delay or reduce their capital expenditures during economic downturns when unemployment is high, as they anticipate lower future demand. Additionally, higher unemployment can indicate an under-utilization of existing capacity. This can lead to lower GFCF, as firms may not need to invest in new capital goods until demand recovers and existing capacity is fully utilized. In some cases, increased capital investment, particularly in automation or technology, can lead to structural unemployment.  \cite{jorgenson_capital_1963} pointed out that in the short term, capital investment might increase unemployment as businesses focus on capital-intensive processes, but in the long term, these investments increase overall productivity and create new employment opportunities. \cite{phelps_money-wage_1968} examined how long-term structural unemployment could be exacerbated by investment in capital goods, especially when new technologies replace jobs. However, automation and robotics can displace jobs in the short term but may lead to productivity gains and job creation in other sectors (see \cite{acemoglu_robots_2020}).

\cite{hall_tax_1967} earlier findings showed that tax policy shapes capital formation by altering the costs and incentives for firms to invest, demonstrating how tax cuts or incentives can amplify corporate investment in capital goods, boosting GFCF. These findings are particularly relevant today as many economies use tax incentives to stimulate investment, showing that while tax incentives can promote GFCF, their design must carefully consider potential unintended outcomes. For instance, when liquidity constraints were reduced and tax incentives were made more attractive, firms increased their domestic capital expenditures \citep{beyer_early_2021,  crawford_effect_2024}. However, in certain cases, firms opted to repatriate foreign cash to invest internationally rather than domestically, suggesting that tax policy design can lead to unintended shifts in investment patterns.

Traditionally, Phillips Curve \citep{phillips_relation_1958} posits that inflation and unemployment are inversely related, and this relationship has broader implications for investment. Higher inflation increases the cost of borrowing due to higher interest rates, which makes capital investments, like GFCF, more expensive for businesses, discouraging them from making long-term investments. Recent research continues to explore how inflation affects business investment by altering the cost of capital and business confidence. Inflation volatility can further erode business confidence, leading to reduced investment in capital-intensive projects. For instance, \cite{pindyck_irreversibility_1990} emphasizes that during periods of uncertainty, particularly related to inflation, firms may choose to delay irreversible investments. Similarly, \cite{montes_effects_2021} shows that economic uncertainty—stemming from both inflation and policy factors—can directly reduce business confidence, which in turn depresses investment levels. Additionally, higher inflation may stymie investments by increasing the uncertainty around future returns, especially when firms cannot predict whether inflationary pressures will subside or persist. Further, \cite{bloom_impact_2009} emphasizes how uncertainty during inflationary periods leads to sharp reductions in investment, as firms prefer to delay capital formation until macroeconomic conditions stabilize. Meanwhile, \cite{smets_shocks_2007} highlight that inflation driven by demand shocks tends to exacerbate business cycle fluctuations, further influencing corporate investment strategies.

A range of control variables was incorporated, including Foreign Direct Investment (FDI), the Economic Policy Uncertainty (EPU) Index, and socioeconomic indicators such as the Human Development Index (HDI) and the Gini Index. FDI plays a crucial role in stimulating GFCF by providing capital, transferring technology, and enhancing managerial expertise, particularly in countries with strong human capital and well-developed financial systems \citep{borensztein_how_1998, blomstrom_multinational_1998, alfaro_fdi_2004}. Thus, it plays a significant role in developing economies with robust institutions, where it boosts capital efficiency and supports infrastructure development \citep{li_impact_2019, nupehewa_more_2022}. However, FDI may also have negative effects, such as trade deficits and reduced domestic investment, potentially leading to economic downturns in some contexts \citep{sabir_institutions_2019, almfraji_foreign_2014}. Uncertainty influences GFCF through business confidence. Firms adjust investments based on expectations of future growth, as shown in studies using Dynamic Stochastic General Equilibrium (DSGE) models, which highlight the role of uncertainty in amplifying economic fluctuations (e.g., \citep{basu_uncertainty_2017, bloom_impact_2009, christiano_risk_2014, christiano_nominal_2005, smets_shocks_2007}). Capital formation supports long-term productivity and human development. Investments in infrastructure and education enhance HDI, creating a feedback loop where improved living standards foster greater economic growth and investment \cite{barro_economic_1989} \cite{pasara_causality_2020}. High income inequality suppresses consumer demand and exacerbates credit constraints for lower-income groups and small businesses, reducing GFCF and limiting long-term GDP \citep{galor_income_1993, stiglitz_price_2015, ostry_ostry_2014}. However, inequality may initially rise with economic growth, but sustained high inequality eventually hinders development by restricting access to education, healthcare, and other public goods, thereby stifling human capital formation \citep{barro_inequality_2000, aghion_chapter_2014}. More equal societies experience longer periods of stable growth and higher GFCF, as wealth distribution allows broader segments of the population to invest in both human and physical capital \citep{berg_redistribution_2018, ostry_inequality_2011}.

\subsection{Econometric models and its limitations}

Linear models have long been the standard in analyzing the impact of macroeconomic factors on corporate investment. However, these models often face limitations, particularly in their inability to fully account for non-linear relationships and their susceptibility to alternative explanations. As a result, the need for non-linear and non-traditional research approaches has become increasingly apparent to enhance the reliability of statistical inference. Despite the progress made, existing knowledge in this area continues to call for further refinement and exploration. Several recent studies have sought to address these gaps by adopting more dynamic econometric approaches, although limitations remain. 

\cite{holm-hadulla_heterogeneity_2021} explored the transmission of monetary policy to corporate decisions using a panel data sample of 10 Eurozone economies. By employing a dynamic framework, they assessed the impact of key macroeconomic variables on external corporate financing. However, while their model accounted for policy shocks, the findings were inconclusive. The effects of monetary tightening depended on the proportion of bonds in corporate financing, highlighting a limitation in the study's ability to generalize across firms with different financing structures from various countries. 

\cite{dao_corporate_2021} extended this analysis to a broader context, examining both developed and emerging market firms across European and non-European countries. Using fixed-effects panel data regression, they studied the relationship between firm-level investment and growth in response to real exchange rate fluctuations, private credit to GDP ratios, and industry-specific labor shares. Although their study offered valuable insights, showing that real currency depreciations significantly impact profits, investment, and asset growth—particularly in firms with higher labor shares—it was constrained by the challenge of mapping certain relationships specific to countries with similar characteristics or level of development, possibly taking into account the dynamic nature of the phenomenon in question by using an endogenous variable as an explanatory variable lagged by one period.

Focusing on the US economy, \cite{bai_effect_2024} investigated the cyclical nature of corporate investment opportunities. Utilizing both cross-sectional and time-series data, they applied a fixed-effects regression model, where investment rates were explained by variables such as GDP percentage change and its interaction with a labor mobility index. Their results demonstrated heterogeneity in the impact of labor mobility on investment, yet the model's reliance on GDP percentage changes and labor mobility indices might have overlooked other potential influences, such as sector-specific shocks or regional disparities, limiting the broader applicability of their findings. 

For developing economies, \cite{jiang_monetary_2024} analyzed Chinese non-financial firms and examined how M2 growth shocks—unexpected changes in money supply—affect corporate investment growth. Using a sample of non-financial firms listed on Chinese stock exchanges, they investigated the impact of unexpected monetary policy changes—specifically, M2 growth shocks—on corporate investment growth. Despite robust findings showing that changes in money supply growth significantly affect firm-level investment growth, based on panel data analysis incorporating both univariate regressions with fixed effects and multivariate regressions with control variables, the study faced limitations. It did not fully account for firms' access to alternative financing channels, which may have mitigated the impact of monetary shocks. Additionally, the focus on Chinese firms raises concerns about the transferability of these results to other developing economies with different monetary policy frameworks.

\cite{hajamini_economic_2018} emphasized the need for cross-sectional analysis to account for the non-linear nature of relationships between variables. For instance, they observed that the relationship between government size and economic growth follows an inverted U-shaped curve, akin to the Barro curve, which can help identify the optimal share of government. In their study, they employed several measures of government size, including the ratio of gross fixed capital formation (GFCF) to gross domestic product (GDP). Analyzing data from 14 developed European countries, they used threshold regression in heterogeneous panel data, as proposed by \cite{hansen_threshold_1999}, to reach their conclusions.

It is important to note, however, that the effects observed by the aforementioned researchers are heterogeneous. Therefore, it would be beneficial to apply additional approaches, such as random effects models or dynamic panel models, to further verify the robustness of the results obtained.

\subsection{Advantages of Random Forest as a Machine Learning Technique}

In this study, we leverage the Random Forest algorithm, an advanced AI-driven ensemble learning method, to model the relationship between GDP growth and corporate investment, focusing on GFCF in firms. Machine learning techniques have recently gained traction in economic forecasting, with Random Forest standing out for its ability to capture complex relationships in economic data \citep{goulet_coulombe_macroeconomy_2024, seeam_comparative_2024, yoon_forecasting_2021} such as economic forecasting, particularly GDP growth prediction \citep{goulet_coulombe_how_2022, maccarrone_gdp_2021, qureshi_forecasting_2021,  yang_machine_2024}.

Introduced by \cite{breiman_random_2001}, Random Forest offers significant advantages over traditional econometric models, particularly in its ability to handle non-linear relationships and interactions between variables. The algorithm constructs multiple decision trees and aggregates their outputs, which allows it to manage high-dimensional datasets, mitigate overfitting, and provide valuable insights into feature importance. In the context of GDP growth and corporate investment, Random Forest has demonstrated superior predictive accuracy compared to traditional models \citep{yoon_forecasting_2021, adriansson_forecasting_2015}. Its ability to capture intricate interactions between macroeconomic variables is especially useful for understanding the complex dynamics of economic systems. Random Forest’s key strengths make it particularly well-suited for predicting GFCF by using panel data because of the following reasons:
\begin{enumerate}
    \item \textbf{Capturing Complex Interactions.} Random Forest models excel at identifying complex, non-linear interactions between variables. In the context of economic forecasting, this capability is particularly valuable as economic relationships are often intricate and not easily captured by linear models \cite{jung_algorithmic_2018}.
    \item \textbf{Improved Predictive Accuracy.} Recent studies have shown that machine learning models, including Random Forest, can outperform traditional econometric models in forecasting GDP growth \cite{richardson_nowcasting_2021}. By applying this approach to corporate investment and GFCF prediction, we extend the frontier of economic forecasting methodologies.
    \item \textbf{Robustness to Outliers and Structural Breaks.} Machine learning models, particularly ensemble methods like Random Forest, have shown resilience in the face of economic crises and structural breaks \cite{jung_algorithmic_2018}. This robustness is crucial when analyzing long-term economic trends that may include periods of significant volatility.
    \item \textbf{Feature Importance Analysis:} Random Forest models provide insights into feature importance, allowing us to identify and rank the most influential factors affecting corporate investment and GFCF. This capability offers valuable insights for policymakers and business leaders \citep{yoon_forecasting_2021}.
\end{enumerate}
Previous studies have applied Random Forest and other machine learning algorithms to economic forecasting, often showing better performance than traditional econometric models, particularly in capturing complex interactions and improving predictive accuracy. However, while Random Forest excels in these areas, it does have limitations in providing direct causal interpretations, which linear models are better suited for.

Although the application of machine learning in economic forecasting is not entirely new \citep{athey_impact_2018, chan_econometrics_2022}, our study introduces a novel approach by specifically applying these techniques to the analysis of corporate investment and GFCF in relation to GDP growth. This integration bridges the gap between traditional economic theory and advanced data science methodologies, offering a more nuanced and potentially more accurate understanding of these critical economic relationships by providing:

\begin{enumerate}
    \item \textbf{Improved Accuracy.} Studies have shown that Random Forest models often outperform both simple autoregressive models (e.g., AR(1)) and more complex linear models in forecasting GDP growth \citep{adriansson_forecasting_2015}. This improved accuracy can be attributed to the algorithm's ability to capture non-linear relationships and interactions among economic indicators.
    \item \textbf{Feature Importance.} Random Forest provides a measure of variable importance, allowing researchers to identify the most influential factors in GDP growth and corporate investment. This feature can offer valuable insights into the drivers of economic performance and guide policy decisions \citep{yoon_forecasting_2021}.
    \item \textbf{Robustness to Outliers.} Economic data often contains outliers, especially during periods of crisis or rapid change. Random Forest's ensemble nature makes it more robust to outliers compared to single-model approaches, potentially improving forecast stability \citep{breiman_random_2001}.
    \item \textbf{Handling of Mixed Data Types.} In economic forecasting, researchers often work with a mix of continuous, categorical, and time-series data. Random Forest can effectively handle these mixed data types without the need for extensive preprocessing \citep{yoon_forecasting_2021}.
\end{enumerate}

The Random Forest model's ability to identify non-linear relationships is particularly valuable in this context, as the impact of GDP growth on corporate investment may vary depending on other economic conditions. For instance, the effect of GDP growth on investment might be amplified during periods of low interest rates or tempered during times of high economic uncertainty. Furthermore, by comparing the performance of Random Forest models with traditional econometric approaches, we can assess the added value of machine learning techniques in economic forecasting. This comparison not only helps validate the robustness of our findings but also contributes to the growing body of literature on the application of advanced analytics in economics \citep{adrian_nicolae_predicting_2023}.

\section{Data and Methodological Approach}

\subsection{Data Sources}
The research sample comprised 32 countries, representing developed and developing economies from each continent (for a full list of countries is available on OSF
\url{https://osf.io/xpyhw/?view_only=aa41c1d52ea54058b5f7d89aa3ea166c}). 
The data for each economy covered the period from 2000 to 2022. Additionally, the countries were categorized based on their membership in international groups or organizations, such as the G7, BRICS, EU-15, and OECD.

The national accounts data, derived from the International Financial Statistics (IFS) database, formed the foundation for cross-country comparisons and were organized using the STAN (STructural ANalysis) framework  \citep{harrigan_estimation_1999, horvati_oecd_2020}. Due to differences in data availability, the study uses an unbalanced panel, meaning the number of observations varied across countries and time periods. Outliers were removed from the dataset to enhance accuracy. Ultimately, a time-series cross-sectional (TSCS) dataset was constructed, consisting of 32 countries over 23 annual periods, yielding approximately 400 observations. The majority of the data was sourced from the International Monetary Fund (IMF) \citep{international_monetary_fund_imf_2024} and World Bank databases \citep{world_bank_world_2024}, while additional control variables were obtained from \cite{economic_policy_uncertainty_economic_2024}  and \cite{human_development_reports_human_2024}  websites.

\begin{table}[h!]
\caption{Descriptive Statistics}
\centering
\begin{tabular}{|l|c|c|c|c|c|c|c|}
\hline
 & Mean & Median & Minimum & Maximum & Std. Dev. & Skewness & Kurtosis \\ \hline
GFCF\_Ratio & 0.2180 & 0.2164 & 0.0515 & 0.5427 & 0.0511 & 0.5737 & 3.2900 \\ \hline
GDP\_Growth & 2.4981 & 2.4259 & -11.1673 & 24.4753 & 3.4266 & -0.0661 & 4.0179 \\ \hline
UnEmpl\_Rate & 7.6046 & 6.2908 & 1.9000 & 29.8800 & 4.8563 & 2.1597 & 5.1403 \\ \hline
TAX & 19.3272 & 20.8857 & 7.7000 & 36.5003 & 6.4674 & -0.0468 & -0.9822 \\ \hline
CPI & 3.3177 & 2.3300 & -4.4800 & 72.3100 & 5.1021 & 7.3497 & 78.5700 \\ \hline
EPU\_Index & 4.7990 & 4.7616 & 3.2281 & 6.4837 & 0.4901 & 0.4741 & 0.4930 \\ \hline
Gini\_Index & 0.3394 & 0.3286 & 0.2380 & 0.5870 & 0.0657 & 1.4441 & 2.2856 \\ \hline
FDI & 4.2858 & 2.4049 & -394.4716 & 234.2487 & 20.9781 & -7.3066 & 200.3755 \\ \hline
HDI & 0.8498 & 0.8880 & 0.4900 & 0.9670 & 0.0963 & -1.3701 & 1.2231 \\ \hline
\end{tabular}
    \begin{tablenotes}
        \footnotesize
        \item Source: Own study.
    \end{tablenotes}
\label{tab:descriptive_stats}
\end{table}

Table~\ref{tab:descriptive_stats} provides information on the statistics describing the variables selected for the study. Due to the nominal values of some variables, constructed as a kind of index, e.g. \textit{GFCF\_Ratio}, \textit{EPU\_Index}, \textit{Gini\_Index} or even \textit{HDI}, the distributions of these variables are closer to normal distributions. The mean and median values are at similar levels, as well as the skewness and kurtosis values at levels close to 0. For other variables, e.g. \textit{GDP\_Growth}, \textit{UnER\_Rate}, \textit{TAX}, \textit{CPI} and \textit{FDI}, relatively large standard deviation values can be seen, as well as high differences in extreme values, i.e. minimum and maximum. Hence, the decision was taken to exclude outliers. At the same time, with regard to the variables \textit{GFCF\_Ratio}, \textit{UnER\_Rate} and \textit{TAX}, data transformation techniques were used to compensate for the perceived certain asymmetry of the distribution. This means that some of the above variables will be expressed in natural logarithm. 

\begin{table}[h!]
\centering
\caption{Correlation Matrix}
\begin{tabular}{|l|c|c|c|c|c|c|c|c|}
\hline
 & GDP\_Growth & UnEmpl\_Rate & TAX & CPI & EPU\_Index & Gini\_Index & FDI & HDI \\ \hline
GDP\_Growth & 1.0000 &  &  &  &  &  &  &  \\ \hline
UnEmpl\_Rate & -0.2129 & 1.0000 &  &  &  &  &  &  \\ \hline
TAX & -0.1583 & 0.0952 & 1.0000 &  &  &  &  &  \\ \hline
CPI & 0.2279 & -0.0584 & -0.2219 & 1.0000 &  &  &  &  \\ \hline
EPU\_Index & -0.1778 & -0.0151 & -0.1057 & -0.0616 & 1.0000 &  &  &  \\ \hline
Gini\_Index & 0.1088 & 0.1151 & -0.5218 & 0.4252 & -0.0167 & 1.0000 &  &  \\ \hline
FDI & 0.0824 & -0.0290 & 0.0881 & -0.0375 & -0.0460 & -0.0846 & 1.0000 &  \\ \hline
HDI & -0.3461 & -0.0392 & 0.3049 & -0.4273 & 0.1602 & -0.5099 & -0.0517 & 1.0000 \\ \hline
\end{tabular}
\begin{tablenotes}
\footnotesize
\item Source: Own study.
\end{tablenotes}
\label{tab:correlation_matrix}
\end{table}

Table~\ref{tab:correlation_matrix} shows the results of examining the associations between the independent variables using correlation coefficients appropriate to the types of variables used. As can be seen from the data presented, the \textit{TAX} variable was quite strongly and negatively correlated with \textit{Gini\_Index} (-0.5218), as well as \textit{CPI} with \textit{HDI} (-0.4273) and \textit{Gini\_Index} with \textit{HDI} (-0.5099), while \textit{CPI} values were moderately strongly and positively correlated with \textit{Gini\_Index} levels (-0.5218). Based on the description of the statistical data carried out, it was possible to select for the models, as potential regressors, variables that were not correlated with each other, i.e. \textit{GDP\_Growth}, \textit{UnEmpl\_Rate}, \textit{TAX} and \textit{CPI}, while the others were treated as control variables or additional instruments that, once verified, will be helpful in testing for causality between variables.

\subsection{Panel Regression Models}

The analysis employs static and dynamic multiple panel regression models to evaluate the influence of economic factors on \textit{GFCF\_Ratio}. To ensure robustness, the analysis incorporates complementary approaches, specifically LSDV (fixed-effects models) and the Generalized Method of Moments (GMM) estimators. In dynamic models, the lagged \textit{GFCF\_Ratio} variable from the previous period is included as an explanatory variable. This choice was supported by the results of the Hausman test, which indicated that the GLS estimator is less consistent than the LSDV estimator \citep{bollen_general_2010, verbeek_guide_2000, jackowicz_political_2014}.

There is a clear rationale for this. The explanatory variables are not constant over time for the countries studied, and it is unlikely that individual effects are uncorrelated with the regressors, as assumed in random effects models. Consequently, a random effects approach could lead to inconsistency due to omitted variable bias \citep{greene_econometric_2000}. Furthermore, robust standard errors were used to construct the panel models. The general form of the regression equation was:
$$
\text{\textit{GFCF\_Ratio}}_{i,t} = f\left( \text{economic\_factors}_{i,t-1}; \text{control\_variables}_{i,t} \right)
$$
where: \textit{GFCF\_Ratio} is a relation of gross fixed capital formation to GDP expressed in natural logarithm; \textit{economic\_factors} are a set of lagged explanatory variables such as \textit{GDP\_Growth} (annual percentage growth rate of GDP at market prices based on constant local currency), \textit{UnEmpl\_Rate} (share of the unemployed in the economically active population); \textit{TAX} (tax revenue referring to compulsory transfers to the central government for public purposes – expressed as a \% of GDP); \textit{CPI} (inflation measured by the Consumer Price Index, which reflects the annual percentage change in the cost of a basket of goods and services for the average consumer); while \textit{control\_variables} are a set of lagged mostly socio-cultural indicators related to the domestic market such as \textit{HDI} (Human Development Index measured a country's overall achievement in its social and economic dimensions, including the health of people, their level of education attainment and their standard of living); \textit{EPU\_Index} (Economic Policy Uncertainty Index); \textit{Gini\_Index} (a measure of the income inequality, the wealth inequality, or the consumption inequality) as well as \textit{FDI} (Foreign Direct Investment as a measure of net inflows of investment to acquire a lasting management interest in an enterprise operating in an economy other than that of the investor). Due to the specific characteristics of the data and the verification of the model using second-order and higher lags for the variables, it was found that the model was of inferior quality or exhibited defects that rendered the estimators inefficient.

The statistical significance of the regression parameters will be assessed using the \textit{t}-test, the standard method for linear regression models. The sign of the statistically significant coefficients will indicate the direction of the relationship. To evaluate the model's fit to the data, we employed the coefficient of determination (\textit{R}-squared). Additionally, to address potential issues of heteroskedasticity and autocorrelation (HAC), we applied Arellano's procedure \cite{arellano_panel_2003}. The Wald statistic was also used to test the joint significance of multiple coefficients.

The literature suggests that the relationships under consideration exhibit a dynamic structure \citep{amighini_fdi_2017, saini_determinants_2018, carvelli_dynamic_2024}. Therefore, we extended our analysis by employing dynamic panel regression models, estimated using the System Generalized Method of Moments (GMM-SYS) procedure, introduced by \cite{blundell_initial_1998}. This approach addresses endogeneity and unobserved heterogeneity by using lagged variables as instruments, effectively instrumenting the endogenous variables with their own lagged values.

Given the issues of endogeneity in our dataset and the similarity between the number of countries and years in our panel (\textit{n} $\approx$ \textit{t}), we opted for the one-step GMM estimator. This approach avoids the potential estimation uncertainty associated with the weighting matrix in the two-step procedure, which can lead to more reliable estimates, particularly in smaller or finite samples \citep{hwang_should_2018}. Additionally, the one-step GMM is often more robust to model misspecification, simplifying the estimation process by using sample moments directly without iterative refinement \citep{imbens_one-step_1997}. In our one-step System GMM estimation, we instrumented the differenced equations using lagged levels and differences of the variables, testing the following relationship:

\[
\text{\textit{GFCF\_Ratio}}_{i,t} = f\left( \text{\textit{GFCF\_Ratio}}_{i,t-1}; \text{economic\_factors}_{i,t-1}; \text{instrumental\_variables}_{i,t-1} \right)
\]

To verify the validity of the instruments, we employed the Sargan test, which examines the validity of over-identifying restrictions by determining whether the instruments used in the model are uncorrelated with the error term. Additionally, we applied the Arellano-Bond tests for first- and second-order autocorrelation \cite{arellano_tests_1991} to check for serial correlation in the idiosyncratic error term. Lastly, the Wald test was used to assess the joint significance of all coefficients and time variables in the model.

\subsection{Random Forest Model}

\begin{figure}[h]
    \centering
    \includegraphics[width=\textwidth]{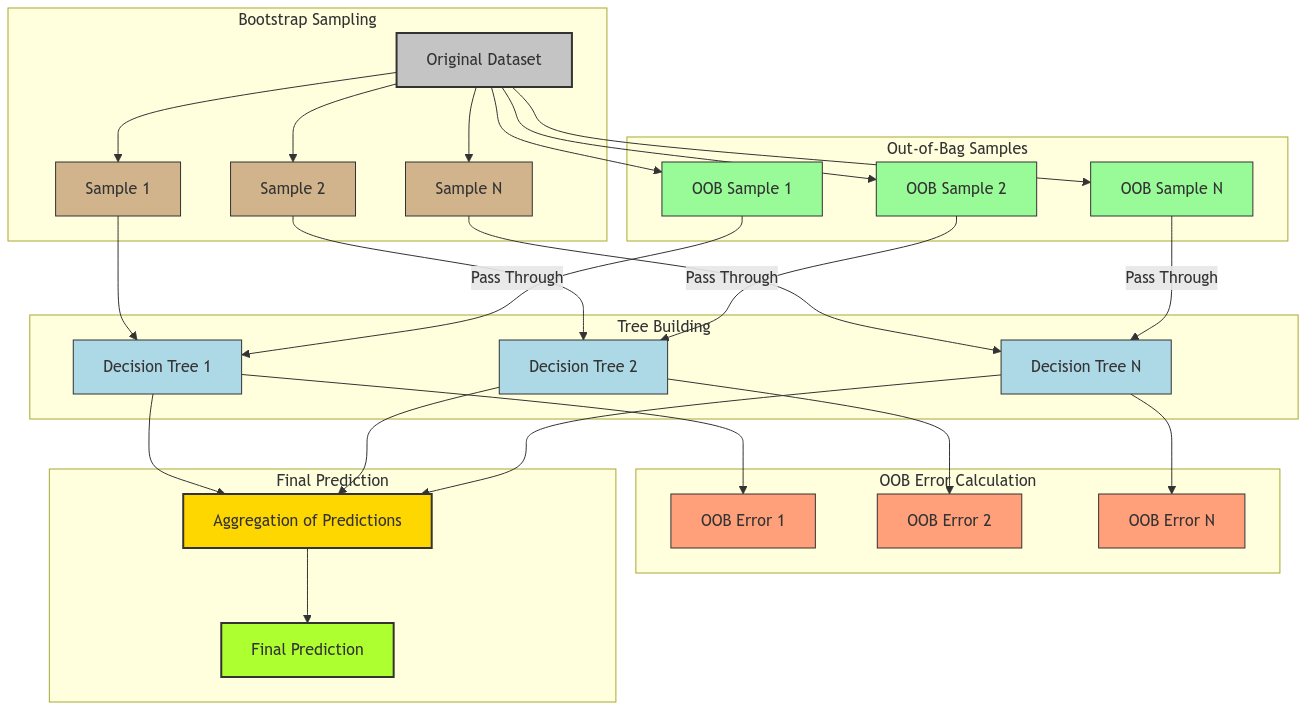}
    \caption{Random Forest Workflow}
    \label{fig:RF}
\end{figure}

In our analysis of the relationship between GDP growth and corporate investment, we employ the Random Forest algorithm. Random Forest is an ensemble learning method that constructs multiple decision trees, each trained on a different bootstrap sample of the original dataset. The algorithm combines the predictions of these individual trees to generate a more accurate and robust final prediction. This technique helps to reduce overfitting and enhances the model's generalization performance. We followed the standard procedure of the Random Forest algorithm, which is outlined as follows (see Fig. \ref{fig:RF}):
\begin{enumerate}
    \item \textbf{Bootstrap Sampling:} For each tree in the forest, a random sample of N records is drawn with replacement from the original dataset.
    \item \textbf{Feature Selection:} At each node of the tree, a random subset of features is selected. The number of features considered at each split is typically the square root of the total number of features for classification tasks, or one-third of the total features for regression tasks.
    \item \textbf{Tree Growing:} The tree is grown by selecting the best split among the randomly chosen features at each node, based on a splitting criterion (e.g., mean squared error for regression).
    \item \textbf{Repeat:} Steps 1-3 are repeated to create multiple trees (typically hundreds or thousands).
    \item \textbf{Aggregation:} For regression tasks, the final prediction is the average of the predictions from all individual trees.
\end{enumerate}

\subsubsection{Panel Data adaptation for Random Forest algorithm}
The Random Forest algorithm, while powerful for many machine learning tasks, is not inherently designed to handle panel data structures, which contain both cross-sectional and temporal dimensions. However, through careful preprocessing and methodological adaptations, we can effectively apply Random Forest to panel data analysis.

In panel data analysis using the Random Forest algorithm, a key preprocessing step involves structuring the dataset to capture temporal dependencies within individual entities. By setting a multi-level index with both entity identifiers (Code) and time periods (Year), we can create a hierarchical data structure that preserves the panel's inherent cross-sectional and time-series characteristics. The subsequent generation of lagged variables through grouped shifting enables the incorporation of past observations as predictive features for each specific entity, addressing the temporal autocorrelation typically encountered in panel data. This methodological approach allows the Random Forest algorithm to leverage historical information while maintaining the unique characteristics of each cross-sectional unit, thereby enhancing the model's ability to capture complex, time-dependent patterns and relationships in the data, despite Random Forest not being natively designed for panel data analysis.

\subsubsection{Out-of-Bag (OOB) Error}

A key feature of Random Forest is its built-in validation mechanism, known as Out-of-Bag (OOB) error estimation. During the bootstrap sampling process, approximately 63.2\% of the original observations are included in the training set for each tree, while the remaining 36.8\% form the OOB set. These OOB samples act as a validation set for each tree, allowing for an unbiased estimate of the model’s performance without the need for a separate test set or cross-validation. The OOB error is computed by aggregating the predictions for each observation using only the trees where that observation was in the OOB set. This method offers a computationally efficient alternative to K-fold cross-validation, as it requires only \(n\) iterations, where \(n\) is the number of trees in the forest.

To enable OOB error estimation in Scikit-learn’s \texttt{RandomForestRegressor} \citep{pedregosa_scikit-learn_2011, scikit-learn_ensembles_2024}, we set the \texttt{oob\_score=True} parameter when initializing the model.

\subsubsection{Evaluating Model Performance: R\textsuperscript{2} Score in Random Forest}

To assess the performance of our model, we use the R² score (R-squared), also known as the coefficient of determination \citep{james_introduction_2021}. The R² score is a statistical measure that indicates the proportion of variance in the dependent variable (corporate investment, in this case) that is explained by the independent variables, such as GDP growth and other macroeconomic factors. It provides a quantitative evaluation of how well the model explains the variability in the data \citep{scikit-learn_r2_score_2024}. For the Random Forest model, the R² score is calculated as follows:

\begin{enumerate} 
\item The algorithm generates predictions for each data point in the test set using the ensemble of decision trees. 
\item The total sum of squares (TSS) is calculated, representing the overall variance in the actual corporate investment values. 
\item The residual sum of squares (RSS) is then computed, which captures the differences between the predicted and actual values. 
\item The $R^2$ score is derived using the formula:  
$$
R^2 = 1 - \frac{RSS}{TSS}
$$
\end{enumerate}
 
The resulting $R^2$ score ranges from 0 to 1, where:

\begin{itemize}
    \item A score of 1 indicates perfect prediction, suggesting that our RandomForest model explains all the variability in corporate investment based on GDP growth and other factors.
    \item A score of 0 implies that the model's predictions are no better than simply using the mean corporate investment value for all predictions.
    \item Scores between 0 and 1 represent the proportion of variance explained by the model. For instance, an $R^2$ of 0.75 suggests that our model accounts for 75\% of the variability in corporate investment.
\end{itemize}

In this study, we compare the $R^2$ scores of the Random Forest model with those of traditional linear regression models. A higher $R^2$ score for the Random Forest model would indicate its superior ability to capture the complex, non-linear relationships between GDP growth and corporate investment \citep{breiman_random_2001}. This comparison allows us to quantify the advantages of employing more advanced machine learning techniques in economic forecasting and policy analysis. However, it is important to emphasize that while a high R² score suggests a strong model fit, it does not imply causality. Therefore, we interpret the results in the broader context of economic theory, alongside other statistical tests, to draw well-rounded conclusions about the relationship between GDP growth and corporate investment.

\subsubsection{Permutation Importance in Random Forest}

Random Forest provides valuable measures of variable importance, helping to identify the most influential predictors in the model. There are two primary methods for calculating variable importance \citep{strobl_bias_2007}:

\begin{itemize}
    \item \textbf{Mean Decrease in Impurity (MDI):} This method quantifies the total reduction in node impurity (e.g., variance reduction in regression) averaged across all trees in the forest. It is available via the \texttt{feature\_importances\_} attribute of the fitted Random Forest regressor.
    
    \item \textbf{Mean Decrease in Accuracy (MDA) or Permutation Importance:} This approach involves randomly permuting the values of each feature and measuring the resulting decrease in model accuracy. A larger decrease indicates greater importance.
\end{itemize}

Permutation importance calculates feature importance by measuring the increase in model prediction error after shuffling a feature's values. If permuting a feature increases the error, the feature is deemed important, as the model relied on it for accurate predictions. Conversely, if the error remains unchanged, the feature is considered unimportant. The permutation importance procedure typically follows five  steps \citep{altmann_permutation_2010}:

\begin{enumerate}
\item \textbf{Establish baseline:} Compute the model's performance (e.g., $R^2$) on the original dataset.
\item \textbf{Permute features:} Randomly shuffle each feature's values, breaking its relationship with the target variable.
\item \textbf{Recalculate performance:} Measure model performance on the permuted dataset.
\item \textbf{Calculate importance:} Feature importance is the difference between the baseline and permuted performance.
\item \textbf{Repeat:} Repeat the process multiple times for each feature to get a distribution of importance scores.
\end{enumerate}

In our study, we use the $permutation\_importance$ function from the sklearn.inspection \citep{scikit-learn_permutation_2024} module. This technique provides a model-agnostic approach to feature importance, which is particularly valuable when dealing with complex dataset. Thus, we can identify the most influential factors in corporate investment, then compare the relative importance of various macroeconomic variables and detect potential redundancies or less important features. Finally, the permutation importance results complement traditional econometric methods and Random Forest's built-in feature importance, offering a comprehensive view of the relationships between GDP growth, corporate investment, and other economic variables.

\subsubsection{p-value Importance in Random Forest}

Unlike linear models where analytical solutions for p-values are available due to well-understood distributional properties, Random Forests lack such theoretical foundations for significance testing. As explained in the paper by  \cite{hapfelmeier_efficient_2023}, the fundamental difference in p-value availability between Random Forests and linear models stems from their underlying mathematical frameworks. Linear models have well-defined parametric distributions that allow for straightforward statistical inference and p-value calculations. However, as explained in the paper, Random Forests are complex machine learning methods where "the distribution of an estimated VIMP under H0 is difficult to assess" and lacks formal distributional assumptions, making traditional p-value computation impossible.

The paper implementation was written in R language. For more verbality and compatibility we had rewritten the p-value importace algorithm in Python language. Our \textit{rfvimptest} algorithm implements a permutation-based approach to test variable importance measures (VIMP) in Random Forests, specifically focusing on permutation VIMP (pVIMP). The algorithm offers several sequential testing methods including Sequential Probability Ratio Test (SPRT), Sequential Approximation to Permutation Test (SAPT), and sequential p-value estimation (PVAL). These methods are designed to reduce computational costs while maintaining statistical validity. The implementation follows the theoretical framework described in the paper, where the null hypothesis H0: pVIMP = 0 is tested against alternatives through permutation testing.

The algorithm incorporates various early stopping criteria based on the theoretical properties discussed in the paper. For SPRT and SAPT, the stopping boundaries are derived from equations (6) and (7) in the paper, while PVAL uses equation (8) for sequential p-value estimation. These stopping rules allow the algorithm to terminate testing before reaching the maximum number of permutations (Mmax) while controlling type-I error probability at the nominal level $\alpha$ and maintaining high power, as demonstrated in the paper's simulation studies.

The implementation includes important practical considerations such as handling missing data, parallel processing for multiple variables, and flexibility in parameter settings. As recommended in the paper, the algorithm allows users to specify the number of trees (ntree) and permutations (nperm) to achieve more stable results, particularly important when dealing with variables whose importance measures are close to the significance threshold. The paper's application studies showed that higher values of ntree especially contribute to more consistent test decisions, which is reflected in the algorithm's parameter options.

\section{Results}

\subsection{To what extent does GDP growth influence Gross Fixed Capital Formation (GFCF), and how does its significance compare to other macroeconomic predictors such as taxation, unemployment, and inflation across emerging (BRICS) and developed (G7, OECD, EU-15) markets?}

This section presents the findings by analyzing the contribution of each explanatory variable in both static and dynamic settings, as summarized in Table~\ref{tab:regr_static} and Table~\ref{tab:regr_dynamic}. Additionally, regression model evaluation metrics are discussed to assess the reliability and fit of the models.

\subsubsection{Static Panel Regression Results}

Table~\ref{tab:regr_static} shows results from static regression models. With regard to the diagnostics of the presented models, it should be emphasized that lower values of coefficients of determination (\textit{R}$^2$) are not unusual in panel data analysis. For instance, in model (OECD), which was constructed using a sample of OECD countries, the goodness-of-fit measure is relatively low, around 10\%. However, this does not impede statistical inference. 
By contrast, models (G7) and (BRICS), which are based on smaller and more homogeneous subsamples, show higher \textit{R}$^2$ values, reaching approximately 46\% and 57\%, respectively. 

In each model presented in Table~\ref{tab:regr_static} , \textit{GDP\_Growth(t-1)} is the regressor with a statistically significant effect on the dependent variable. 
For the \textbf{G7}, the coefficient is \( 0.0112 \) with its significance score of \( p < 0.01 \). In the BRICS, the coefficient is slightly higher at \( 0.0128 \), also significant at the 5\% level. For the EU-15, GDP growth contributes \( 0.0088 \), significant at the 1\% level. In the OECD, the coefficient is \( 0.0093 \) and remains significant at the 1\% level. 
Across all research samples, a one-unit increase in \textit{GDP\_Growth} from the previous period led to an average increase of 1\% in the \textit{GFCF\_Ratio}, holding all other variables constant. This suggests that \textit{GDP\_Growth} is both a statistically and economically significant determinant of GFCF. 

In contrast, two other independent variables, \textit{UnEmpl\_Rate} and \textit{TAX}, act mostly as destimulants. In the first case, 
for the G7, the coefficient next to the \textit{LN\_UnEmpl\_Rate(t-1)} is \( -0.1354 \), significant at the 1\% level. In the BRICS, the coefficient is \( -0.2250 \), the largest magnitude, also significant at the 1\% level. For the EU-15, the coefficient is \( -0.1280 \), significant at the 1\% level. In the OECD, unemployment shows a smaller magnitude with \( -0.0349 \) and is significant at the 10\% level. 
In the second case, the effect of \textit{LN\_TAX(t-1)} varies across regions. For the G7, taxation strongly reduces GFCF, with a coefficient of \( -0.4551 \), significant at the 1\% level. In the EU-15, the coefficient is \( -0.2602 \) and significant at the 5\% level. By contrast, in the BRICS, taxation has a positive coefficient of \( 0.4077 \), significant at the 10\% level. In the OECD, taxation negatively affects GFCF (\( -0.2330 \)), significant at the 1\% level. 
In general, an increase in the unemployment rate (share of the unemployed in the economically active population) and higher compulsory tax transfers to the central government from the previous period are associated with a decrease in the \textit{GFCF\_Ratio} in the subsequent period. 

The last explanatory variable, the consumer price index (CPI), shows limited significance across regions. 
The \textit{CPI(t-1)} for the OECD, inflation has a positive coefficient (\( 0.0123 \)), significant at the 1\% level. In other regions (G7, BRICS, and EU-15), the variable remains insignificant. 
The statistically significant results in only one model (OECD) suggest that these findings should be interpreted with caution.

\begin{table}
    \centering
    \caption{Static Panel Regression Models}
    \begin{filecontents*}{tables_DF/Table_3.csv}
FEM;G7;;BRICS;;EU-15;;OECD;;
;(1);;(2);;(3);;(4);;
const;-0,0698;;-2,1708;***;-0,5296;;-0,8305;***;
;(0.2361);;(0.4658);;(0.3290);;(0.2287);;
GDP\_Growth(t-1);0,0112;***;0,0128;**;0,0088;***;0,0093;***;
;(0.0028);;(0.0053);;(0.0024);;(0.0024);;
LN\_UnEmpl\_Rate(t-1);-0,1354;***;-0,2250;***;-0,1280;***;-0,0349;*;
;(0.0176);;(0.0638);;(0.0198);;(0.0181);;
LN\_TAX(t-1);-0,4551;***;0,4077;*;-0,2602;**;-0,2330;***;
;(0.0822);;(0.2082);;(0.1051);;(0.0748);;
CPI(t-1);-0,0004;;-0,0107;;0,0055;;0,0123;***;
;(0.0048);;(0.0075);;(0.0048);;(0.0042);;
Observations;123;;31;;237;;388;;
R-squared;0,4585;;0,5720;;0,2733;;0,0992;;
Adj. R-squared;0,4402;;0,5061;;0,2608;;0,0898;;
F-statistic;23,7084;***;7,6840;***;20,5922;***;9,8855;***;
\end{filecontents*}
\csvautotabular[separator=semicolon]{tables_DF/Table_3.csv}
    \label{tab:regr_static}
    \begin{tablenotes}
        \footnotesize
        \item Note: The table presents the results of the estimations for the random effects model. For brevity, estimates for year dummies and control variables are omitted. ***, **, and * denote significance at the 1\%, 5\%, and 10\% levels, respectively. Robust standard errors are shown in parentheses.
        \item Source: Own study.
    \end{tablenotes}
\end{table}

\subsubsection{Dynamic Panel Regression Results}

In Table~\ref{tab:regr_dynamic}, which presents findings from the estimation of dynamic panel models, it is notable that the Sargan test results indicate no grounds to reject the hypothesis of over-identifying restrictions, confirming the validity of the instruments used in the models. 
As we can see in Table~\ref{tab:regr_dynamic}, the primary determinant of \textit{GFCF\_Ratio} is the previous period's GFCF values, which are statistically significant across all four models, underscoring the long-term nature of corporate investment and its reliance on prior expenditures as well as economic conditions. 
The lagged \textit{LN\_GFCF\_Ratio(t-1)} exhibits the highest coefficient in the OECD (\( 0.9856, p < 0.01 \)), reflecting strong investment inertia in advanced and economically diverse economies. In the G7, a significant coefficient of \( 0.9595 \) (\( p < 0.01 \)) highlights stable and persistent investment patterns in mature markets. BRICS, though slightly lower, still shows a highly significant coefficient (\( 0.9238, p < 0.01 \)), but due to some anomalies (the statistical significance of AR(2)), it was not possible to include all explanatory variables into the model at the same time. Similarly, the EU-15 reports a coefficient of \( 0.9594 \) (\( p < 0.01 \)), affirming the persistence of GFCF in economically integrated regions.

Additionally, other macroeconomic indicators partially explain the level of \textit{GFCF\_Ratio}. For instance, 
the coefficient next to \textit{GDP\_Growth(t-1)} for the G7 is \( 0.0121 \) and remains highly significant (\( p < 0.01 \)). In the EU-15, the effect is smaller, with a coefficient of \( 0.0068 \) (\( p < 0.01 \)), indicating a weaker but still significant influence. With regard to the other models, this variable was not statistically significant. 
Hence, the economic growth has a statistically significant effect on the dependent variable, particularly for larger economies (G7) or economically and geographically integrated regions (EU-15) but diminishes in magnitude compared to the static models. 

In the dynamic approach, no significant effect of unemployment rate or taxation on GFCF is observed. So was another of the variables, i.e. inflation. Although in the case of the EU-15, the coefficient next to \textit{CPI(t-1)} is negative (\( -0.0054, p < 0.01 \)), indicating that price instability may deter GFCF. However it should be treated with caution.

\begin{table}
    \centering
    \caption{Dynamic Panel Regression Models}
    \begin{filecontents*}{tables_DF/Table_4.csv}
GMM-SYS;G7;;BRICS;;EU-15;;OECD;;
;(5);;(6);;(7);;(8);;
LN\_GFCF\_Ratio(t-1);0,9595;***;0,9238;***;0,9594;***;0,9856;***;
;(0.0180);;(0.0531);;(0.0217);;(0.0208);;
GDP\_Growth(t-1);0,0121;***;;;0,0068;***;0,0027;;
;(0.0019);;;;(0.0023);;(0.0024);;
LN\_UnEmpl\_Rate(t-1);0,0157;;-0,0112;;0,0105;;0,0085;;
;(0.0122);;(0.0158);;(0.0072);;(0.0055);;
LN\_TAX(t-1);-0,0017;;;;0,0037;;0,0004;;
;(0.0057);;;;(0.0056);;(0.0074);;
CPI(t-1);-0,0024;;;;-0,0054;***;-0,0041;;
;(0.0022);;;;(0.0020);;(0.0037);;
Observations;123;;31;;237;;388;;
Sargan test;7;;5;;14;;25;;
AR(1);-2,2742;**;-1,7185;*;-1,9348;*;-2,0901;**;
AR(2);1,5965;;-1,9249;*;0,4018;;-0,1545;;
\end{filecontents*}
\csvautotabular[separator=semicolon]{tables_DF/Table_4.csv}
    \label{tab:regr_dynamic}
    \begin{tablenotes}
        \footnotesize
        \item Note: The table reports estimation results for the one-step System GMM model. For brevity, estimates for year dummies and instrument variables are omitted. ***, **, and * denote significance at the 1\%, 5\%, and 10\% levels, respectively. Robust standard errors are shown in parentheses.
        \item Source: Own study.
    \end{tablenotes}
\end{table}

\subsubsection{Panel Regression Evaluation Metrics}\label{LR_Performance_Metrics}

The evaluation metrics of the static panel regression models is assessed using \( R^2 \), Adjusted \( R^2 \), and the \textit{F}-statistic across all groups. For the G7, the model explains approximately \( 45.85\% \) of the variation in the dependent variable \( \text{\textit{GFCF\_Ratio}} \), with an Adjusted \( R^2 \) of \( 0.4402 \), reflecting a good model fit after accounting for the number of predictors. The \textit{F}-statistic is \( 23.7084 \), highly significant (\( p < 0.01 \)), confirming the joint significance of the model's variables. The BRICS model achieves the highest \( R^2 \) value among all regions at \( 0.5720 \), indicating that \( 57.20\% \) of the variation in \( \text{\textit{GFCF\_Ratio}} \) is explained by the independent variables. Despite the smallest sample size (\( N = 31 \)), the Adjusted \( R^2 \) is \( 0.5061 \), confirming a strong model fit. The \textit{F}-statistic is \( 7.6840 \), significant at \( p < 0.01 \), validating the robustness of the model. For the EU-15 region, the \( R^2 \) is \( 0.2733 \), indicating that \( 27.33\% \) of the variation in \( \text{\textit{GFCF\_Ratio}} \) is explained by the model. Although this value is relatively low, it is not unusual for panel data models where heterogeneity and unobserved factors often influence the unexplained variation. The Adjusted \( R^2 \) is \( 0.2608 \), lower compared to G7 and BRICS but still acceptable for panel data analysis. The \textit{F}-statistic is \( 20.5922 \), highly significant (\( p < 0.01 \)), ensuring the overall significance of the model. The OECD model exhibits the lowest \( R^2 \) value at \( 0.0992 \), meaning that only \( 9.92\% \) of the variation in \( \text{\textit{GFCF\_Ratio}} \) is explained by the model. This result is particularly notable given the OECD sample has the largest number of observations (\( N = 388 \)). The relatively low \( R^2 \) may be attributed to the economic diversity and heterogeneity within the OECD countries, making it more challenging for the model to capture variations in \textit{\( \text{\textit{GFCF\_Ratio}} \)}. The Adjusted \( R^2 \) is \( 0.0898 \), further reflecting the limited explanatory power, but it should be explained that some explanatory variables are indices with low nominal values or limited ranges, which can naturally reduce the \textit{R}$^2$ or Adjusted \( R^2 \) without compromising the validity of the model  \citep{cox_comment_1992}. Consequently, it should not be viewed as indicative of poor model fit in panel data contexts. Nevertheless, the \textit{F}-statistic of \( 9.8855 \) remains significant at \( p < 0.01 \), confirming that the independent variables jointly contribute to explaining the dependent variable.

The dynamic models demonstrate superior performance compared to their static counterparts, as evidenced by the diagnostic tests. The Sargan test confirms the validity of the instruments used across all models, as the null hypothesis of over-identifying restrictions is not rejected. For the G7, the Sargan statistic indicates valid instruments for this advanced economy group, with no issues detected. In the BRICS model, the Sargan statistic of 5 confirms instrument validity despite the smaller sample size. For the EU-15, the Sargan statistic of 14 is a consequence of the larger sample size but does not indicate any problems with instrument validity. Similarly, in the OECD model, the Sargan statistic of 25, while higher due to the complex economic structure of the group, confirms that the instruments used remain valid. The AR(1) and AR(2) tests further validate the models. AR(1) results are significant across all regions, confirming first-order serial correlation. Importantly, the AR(2) test results are not significant across all models, indicating the absence of second-order serial correlation and confirming the validity of the dynamic panel regression models. For the G7, the AR(1) test statistic is \( -2.2742 \) (\( p < 0.05 \)), while the AR(2) statistic is \( 1.5965 \), showing no evidence of second-order correlation. 
An anomaly arises in the BRICS model, where the 2nd-order autocorrelation test, AR(2), takes statistically significant values (\( -1.9249 \), \( p < 0.1 \)). This may be attributed to the small sample size (the BRICS subsample included only 5 countries), high correlation among the independent variables, or the potential endogeneity within the analyzed relationship. Specifically in the model of BRICS, it was challenging to simultaneously match variables like \textit{GDP\_Growth}, \textit{TAX}, and \textit{CPI}.
For the EU-15, AR(1) is \( -1.9348 \) (\( p < 0.05 \)), and AR(2) is \( 0.4018 \), confirming no significant second-order correlation. Similarly, for the OECD, AR(1) is \( -2.0901 \) (\( p < 0.05 \)), and AR(2) is \( -0.1545 \), further validating the robustness of the instruments and the absence of problematic serial correlation. The dynamic panel regression results confirm the dominance of lagged GFCF values across all regions, reflecting the persistence of investment patterns. GDP growth remains significant but declines in magnitude compared to static models. Other macroeconomic indicators, such as unemployment and inflation, exhibit weaker effects. 

\subsection{Which macroeconomic factors, including GDP growth, taxation, unemployment, and lagged GFCF, emerge as the most significant predictors of GFCF according to variable importance in Random Forest models?}

\begin{figure}[h]
    \centering
    \begin{subfigure}{0.45\textwidth}
        \centering
        \includegraphics[width=\linewidth]{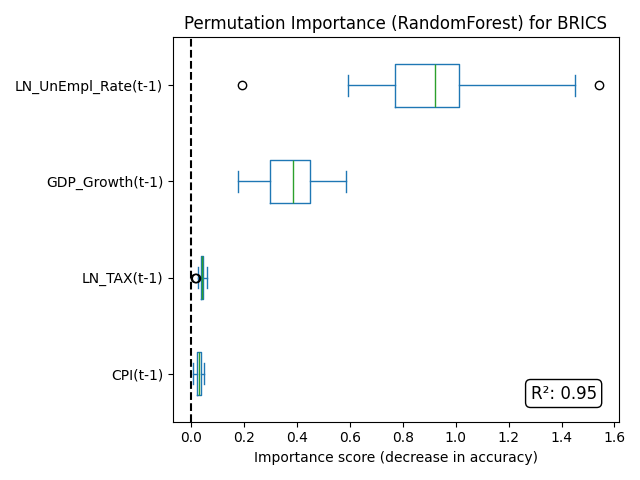}
    \end{subfigure}
    \hfill
    \begin{subfigure}{0.45\textwidth}
        \centering
        \includegraphics[width=\linewidth]{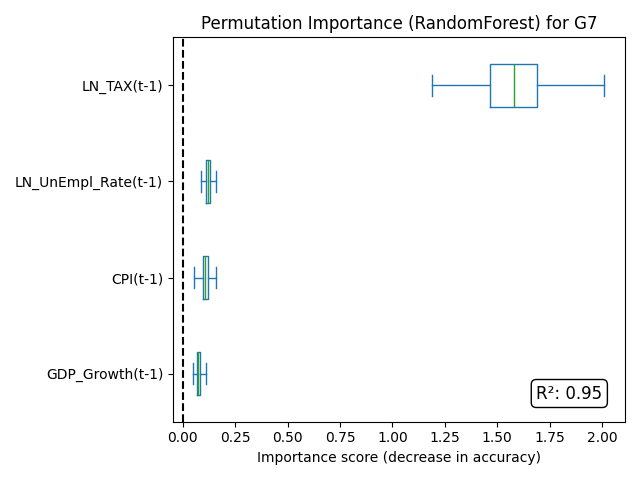}
    \end{subfigure}
    
    \vskip\baselineskip
    \begin{subfigure}{0.45\textwidth}
        \centering
        \includegraphics[width=\linewidth]{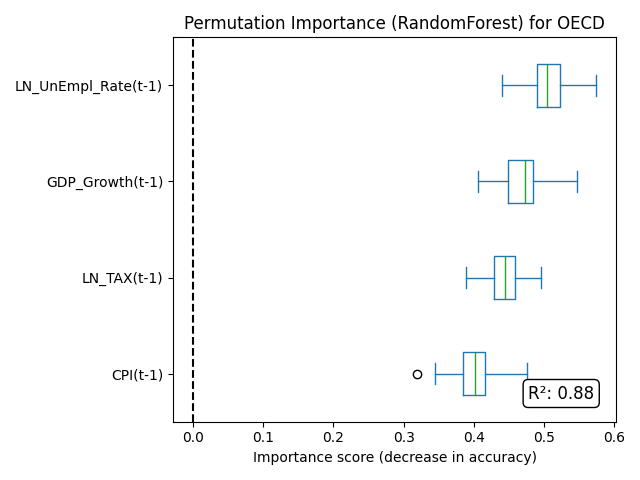}
    \end{subfigure}
    \hfill
    \begin{subfigure}{0.45\textwidth}
        \centering
        \includegraphics[width=\linewidth]{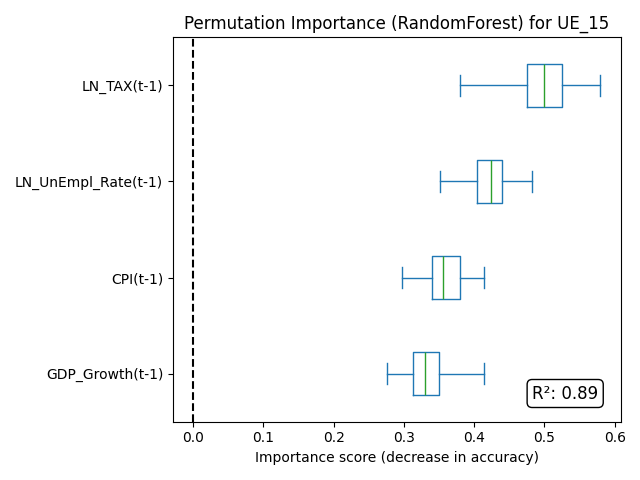}
    \end{subfigure}

    \caption{Permutation importance of Random Forest algorithm for \textbf{Static Panel Regression}.}
    \label{fig:2x2_plots_PI_no_lag}
\end{figure}

\begin{figure}[h]
    \centering
    \begin{subfigure}{0.45\textwidth}
        \centering
        \includegraphics[width=\linewidth]{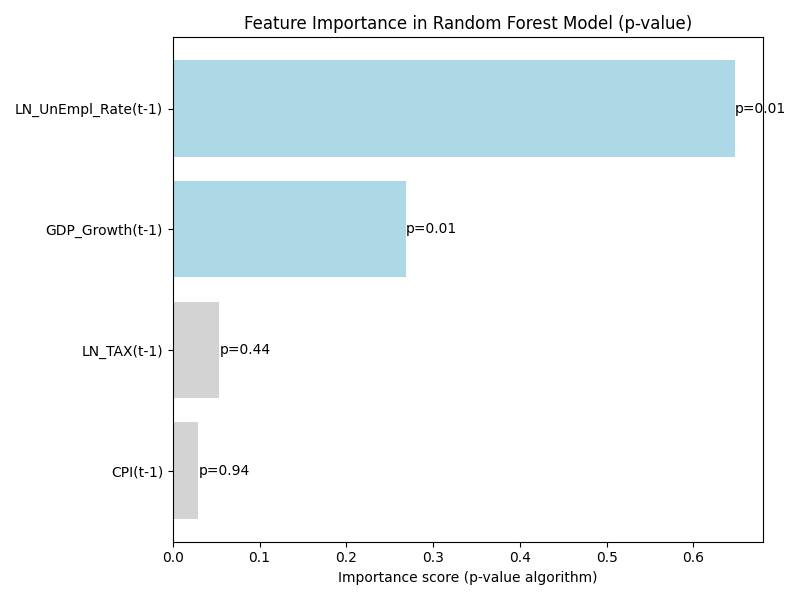}
    \end{subfigure}
    \hfill
    \begin{subfigure}{0.45\textwidth}
        \centering
        \includegraphics[width=\linewidth]{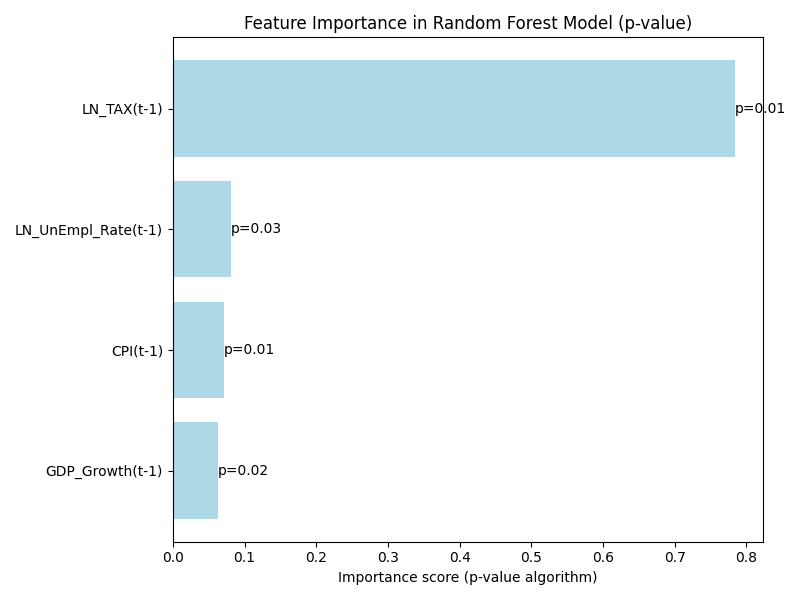}
    \end{subfigure}
    
    \vskip\baselineskip
    \begin{subfigure}{0.45\textwidth}
        \centering
        \includegraphics[width=\linewidth]{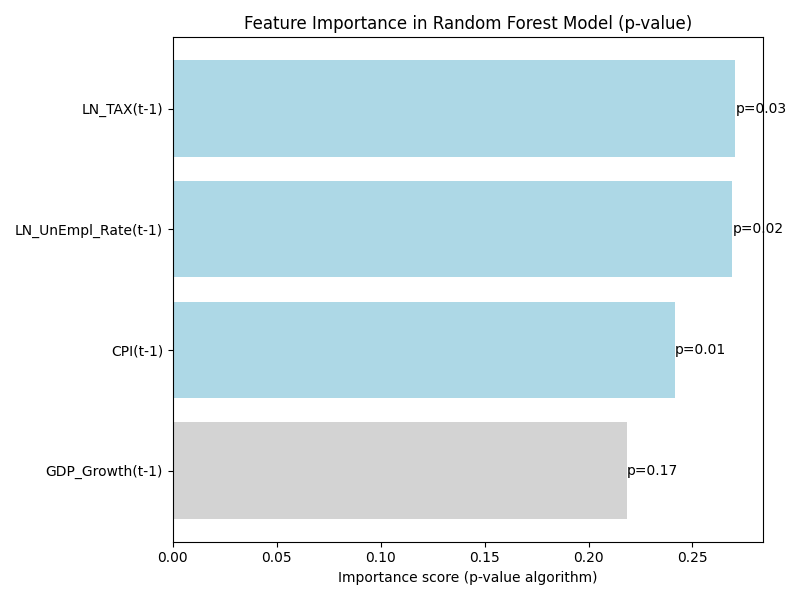}
    \end{subfigure}
    \hfill
    \begin{subfigure}{0.45\textwidth}
        \centering
        \includegraphics[width=\linewidth]{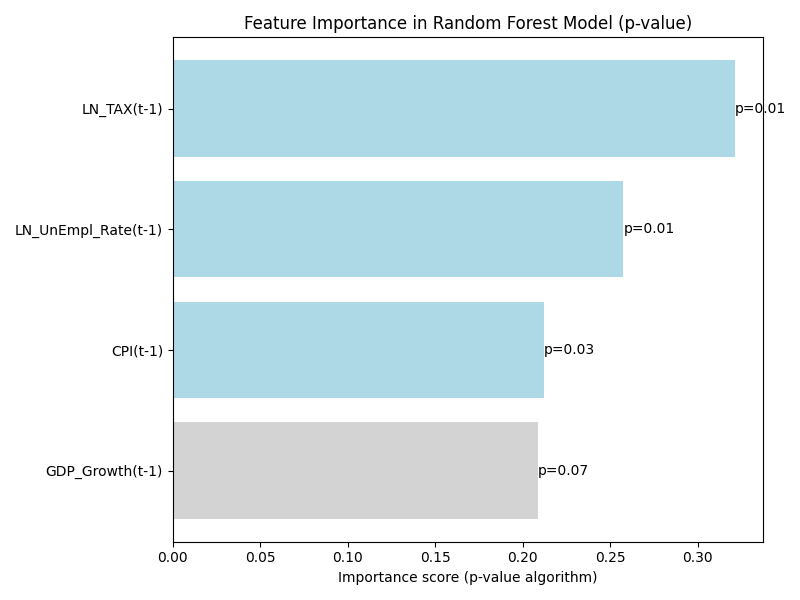}
    \end{subfigure}

    \caption{P-value importance of Random Forest algorithm for \textbf{Static Panel Regression}. In gray are variables/features with p-value > 0.05. Importance algorithm is based on  sequential p-value estimation \cite{hapfelmeier_efficient_2023}}
    \label{fig:2x2_plots_Pvalue_no_lag}
\end{figure}

\begin{figure}[h]
    \centering
    \begin{subfigure}{0.45\textwidth}
        \centering
        \includegraphics[width=\linewidth]{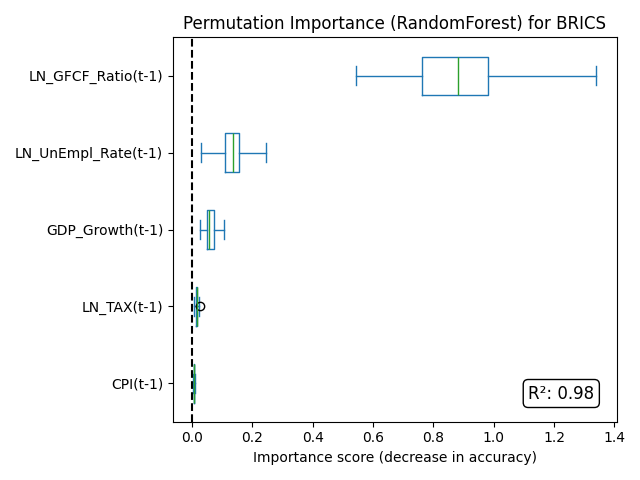}
    \end{subfigure}
    \hfill
    \begin{subfigure}{0.45\textwidth}
        \centering
        \includegraphics[width=\linewidth]{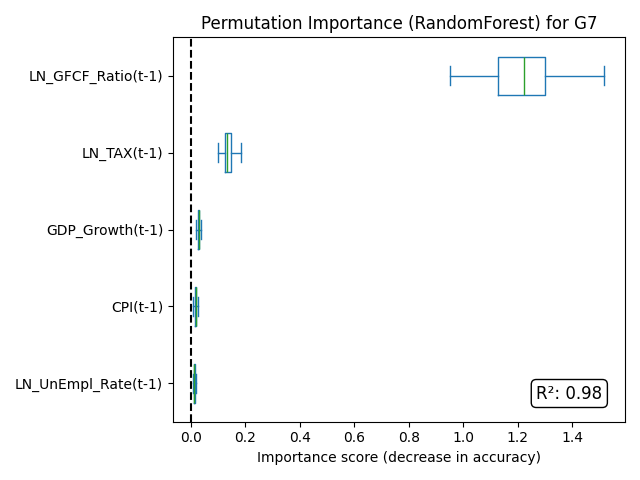}
    \end{subfigure}
    
    \vskip\baselineskip
    \begin{subfigure}{0.45\textwidth}
        \centering
        \includegraphics[width=\linewidth]{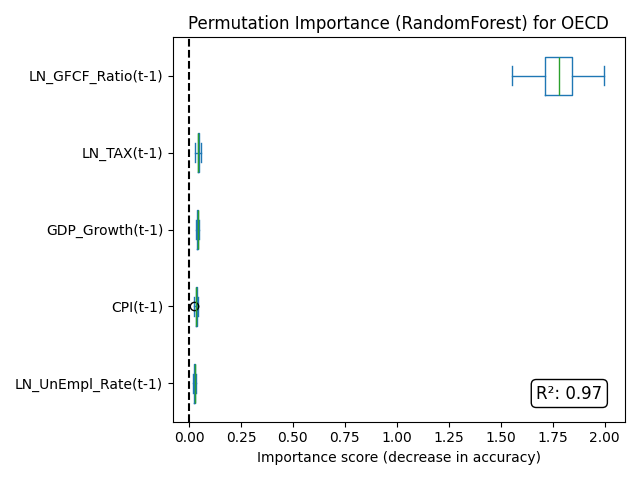}
    \end{subfigure}
    \hfill
    \begin{subfigure}{0.45\textwidth}
        \centering
        \includegraphics[width=\linewidth]{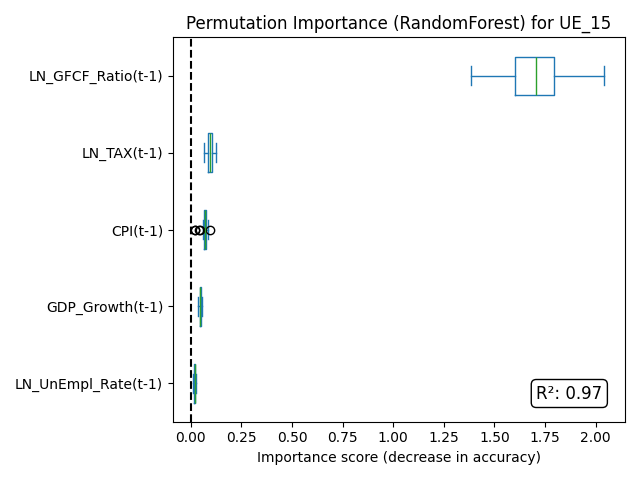}
    \end{subfigure}

    \caption{Permutation importance form Random Forest algorithm for \textbf{Dynamic Panel Regression}.}
    \label{fig:2x2_plots_PI_lag}
\end{figure}

\begin{figure}[h]
    \centering
    \begin{subfigure}{0.45\textwidth}
        \centering
        \includegraphics[width=\linewidth]{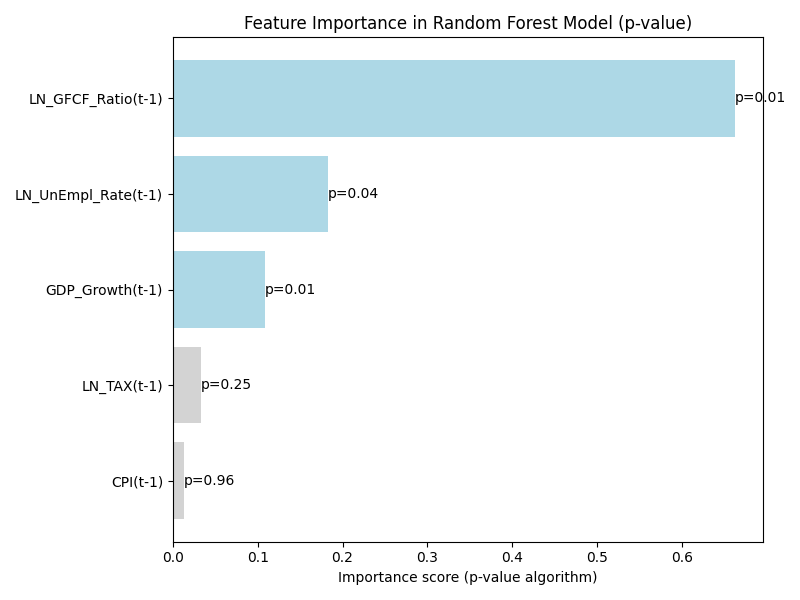}
    \end{subfigure}
    \hfill
    \begin{subfigure}{0.45\textwidth}
        \centering
        \includegraphics[width=\linewidth]{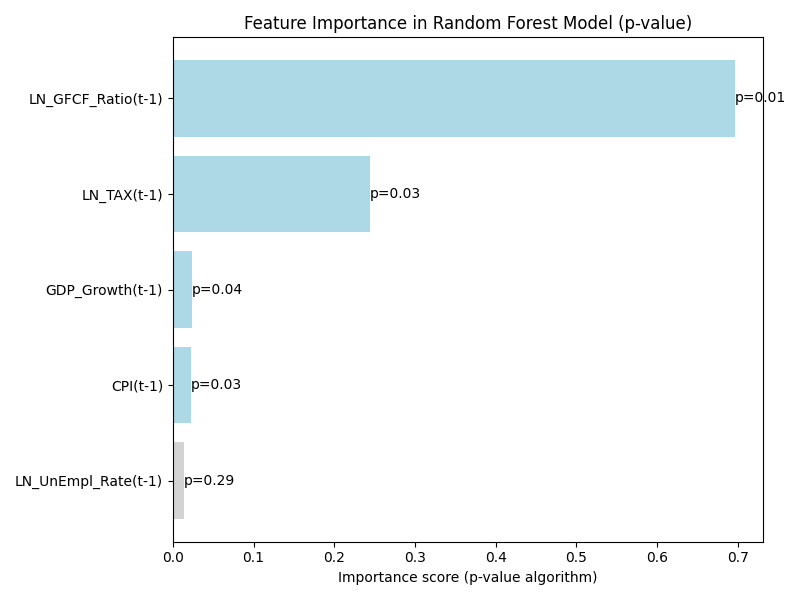}
    \end{subfigure}
    
    \vskip\baselineskip
    \begin{subfigure}{0.45\textwidth}
        \centering
        \includegraphics[width=\linewidth]{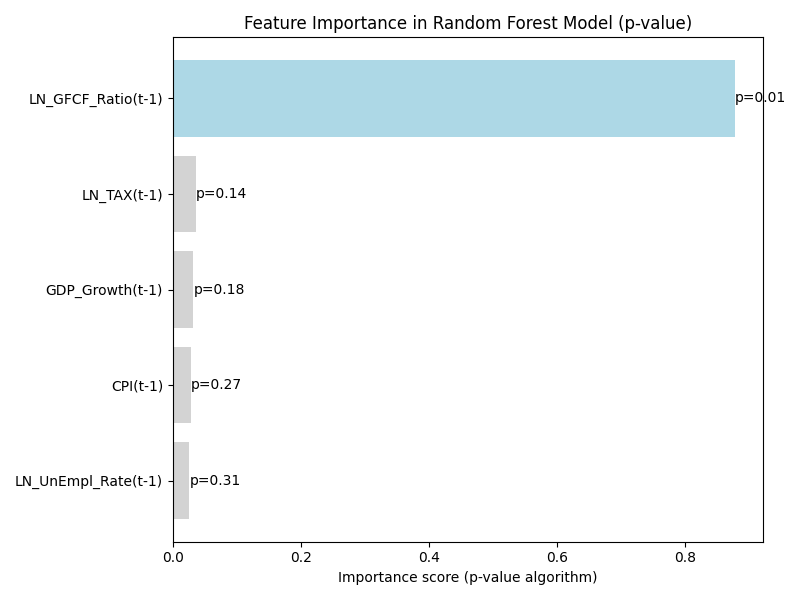}
    \end{subfigure}
    \hfill
    \begin{subfigure}{0.45\textwidth}
        \centering
        \includegraphics[width=\linewidth]{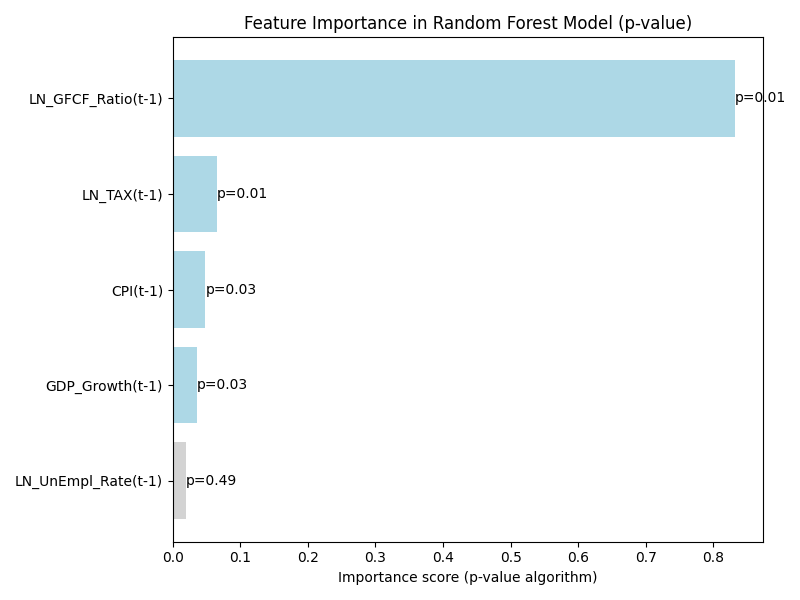}
    \end{subfigure}

    \caption{P-value importance form Random Forest algorithm for \textbf{Dynamic Panel Regression}. In gray are variables/features with p-value > 0.05. Importance algorithm is based on  sequential p-value estimation \cite{hapfelmeier_efficient_2023}}
    \label{fig:2x2_plots_Pvalue_lag}
\end{figure}

This section presents the findings by analyzing the relative importance of each explanatory variable in both static and dynamic settings, as summarized in Table~\ref{tab:rf_static} and Table~\ref{tab:rf_dynamic}. Additionally, model evaluation metrics, including \( R^2 \), Adjusted \( R^2 \), and Mean Squared Error (MSE), are discussed to assess the accuracy and reliability of the Random Forest models. The results are further illustrated in Figures~\ref{fig:2x2_plots_PI_no_lag}, \ref{fig:2x2_plots_Pvalue_no_lag}, \ref{fig:2x2_plots_PI_lag}, and \ref{fig:2x2_plots_Pvalue_lag}, which provide insights into the permutation importance score and statistical significance of each variable.

\subsubsection{Random Forest in Static Settings}

GDP growth (\( \text{\textit{GDP\_Growth}(t-1)} \)) consistently demonstrates significance but shows regional variation in importance score. In OECD, GDP growth achieves the highest importance score (\( 0.4700 \), \( p < 0.01 \)), underscoring its role in explaining \( \text{\textit{GFCF\_Ratio}} \) in advanced and economically diverse economies. In BRICS, GDP growth importance score is \( 0.3779 \) (\( p < 0.01 \)), reflecting the strong reliance of emerging economies on short-term economic expansion to drive investments. For the EU-15, the importance score is moderate (\( 0.3319 \), \( p < 0.05 \)), highlighting economic growth’s contribution to investment dynamics in integrated economies. GDP growth in G7 has the lowest importance (\( 0.0753 \), \( p > 0.05 \)), suggesting a reduced sensitivity of mature economies to short-term growth compared to fiscal or structural determinants.

The unemployment rate (\( \text{\textit{LN\_UnEmpl\_Rate}(t-1)} \)) emerges as the dominant predictor in BRICS, with a high importance score of \( 0.9080 \) (\( p < 0.01 \)), indicating the significant impact of labor market conditions on investments in emerging economies. In OECD, unemployment retains moderate importance (\( 0.5064 \), \( p < 0.05 \)), highlighting its role in varied economies' investment decisions. For EU-15, importance score is \( 0.4213 \) (\( p < 0.05 \)), reinforcing its relevance to investment behavior in integrated markets. In G7, unemployment has a smaller importance score (\( 0.1202 \), \( p < 0.05 \)), indicating limited influence in mature labor markets.

Taxation (\( \text{\textit{LN\_TAX}(t-1)} \)) emerges as a critical determinant, particularly in advanced economies. In G7, taxation has the highest importance score across all variables (\( 1.5784 \), \( p < 0.01 \)), highlighting the significant role of fiscal policy in influencing fixed capital investments. Similarly, the EU-15 taxation remains highly relevant (\( 0.4988 \), \( p < 0.05 \)), reflecting diverse fiscal policies and their impact on investment behavior. The OECD importance score is \( 0.4437 \) (\( p < 0.05 \)), further emphasizing the importance of tax incentives in advanced and heterogeneous economies.
Whereas, in BRICS taxation is negligible (\( 0.0388 \), \( p > 0.1 \)), aligning with the less structured fiscal frameworks and limited reliance on tax-based investment strategies in emerging markets.

The Consumer Price Index (\( CPI(t-1) \)) demonstrates moderate importance in OECD (\( 0.4002 \) (\( p < 0.05 \)), and EU-15 (\( 0.3586 \) (\( p < 0.05 \)). The minimal importance in G7 (\( 0.1055 \), \( p < 0.01 \)) and the negligible importance in BRICS (\( 0.0281 \), \( p > 0.05 \)) suggests that inflation is a less direct determinant of GFCF in these regions.

\begin{table}
    \centering
    \caption{RandomForest: \textbf{Static Panel Regression} Models. 
    }
    \begin{filecontents*}{RF_no_lag/RF_result_summary_no_lag.csv}
Metric,G7,BRICS,EU-15,OECD
GDP\_Growth(t-1),0.0753  **,0.3779 ***,0.3319   *,0.4700 
,(0.0127),(0.0912),(0.0275),(0.0295)
LN\_UnEmpl\_Rate(t-1),0.1202  **,0.9080 ***,0.4213 ***,0.5064  **
,(0.0143),(0.2015),(0.0298),(0.0265)
LN\_TAX(t-1),1.5784 ***,0.0388 ,0.4988 ***,0.4437  **
,(0.1675),(0.0073),(0.0378),(0.0209)
CPI(t-1),0.1055 ***,0.0281 ,0.3586  **,0.4002 ***
,(0.0187),(0.0094),(0.0271),(0.0256)
Observations,123,28,233,376
MSE,0.0002,0.0013,0.0018,0.0022
R-squared,0.9841,0.9533,0.8887,0.8840
Adj. R-squared,0.9835,0.9452,0.8867,0.8827
F-statistic,1355.8744 ***,99.2338 ***,272.3469 ***,410.8423 ***
\end{filecontents*}
\csvautotabular[separator=comma]{RF_no_lag/RF_result_summary_no_lag.csv}
    \label{tab:rf_static}
    \begin{tablenotes}
        \footnotesize
        \item Note:
        Value is permutation importance score with standard deviation in brackets below. The p-value significance code of variable importance is presented, and the p-value is based on sequential p-value estimation \cite{hapfelmeier_efficient_2023}.
        \\
        For brevity, estimates for year dummies and instrument variables are omitted. ***, **, and * denote significance at the 1\%, 5\%, and 10\% levels, respectively. Robust standard errors are shown in parentheses.
        \item Source: Own study.
    \end{tablenotes}
\end{table}

\subsubsection{Random Forest in Dynamic Settings}

Lagged GFCF ratio (\( LN\_GFCF\_Ratio(t-1) \)) is the most important predictor across all regions, confirming the path-dependent nature of fixed capital formation. In the OECD, it exhibits the highest importance score of \( 1.7738 \) (\( p < 0.01 \)), indicating a strong reliance on prior investments. The EU-15 follows closely with a score of \( 1.6970 \) (\( p < 0.01 \)), reflecting the importance of historical capital stock in economically integrated regions. In G7, the lagged GFCF ratio retains strong significance with a score of \( 1.2246 \) (\( p < 0.01 \)), highlighting stability in advanced economies. For BRICS, the importance score is lower at \( 0.8789 \) (\( p < 0.01 \)), but it remains the dominant determinant.

The importance of GDP growth (\( GDP\_Growth(t-1) \)) diminishes across all regions, with scores below \( 0.1 \). This suggests that its effect is largely mediated by lagged GFCF values. BRICS demonstrates the highest sensitivity to GDP growth with an importance score of \( 0.0604 \) (\( p < 0.05 \)). In the EU-15, the score is \( 0.0461 \), but it is not statistically significant (\( p > 0.05 \)), reflecting a weaker influence. Similarly, OECD (\( 0.0414, p > 0.05 \)) and G7 (\( 0.0280, p > 0.05 \)) show minimal importance, highlighting that GDP growth has a limited direct impact in advanced economies under dynamic settings.

The unemployment rate  (\( LN\_UnEmpl\_Rate(t-1) \)) retains partial importance in BRICS, where it has a score of \( 0.1351 \) (\( p < 0.05 \)). In contrast, the importance scores for the EU-15 (\( 0.0191 \)), OECD (\( 0.0271 \)), and G7 (\( 0.0123 \)) remain minimal and statistically insignificant (\( p > 0.05 \)).

Taxation (\( LN\_TAX(t-1) \)) plays a moderate role in G7 (\( 0.1357, p < 0.05 \)) and EU-15 (\( 0.0956, p < 0.05 \)), reflecting the reliance on fiscal policies and incentives in these regions. In BRICS (\( 0.0153, p > 0.05 \)) and OECD (\( 0.0448, p > 0.05 \)), the importance of taxation is lower and statistically insignificant, suggesting that tax policies have a reduced role in influencing GFCF in these regions.

The Consumer Price Index (\( CPI(t-1) \)) demonstrates minimal importance across all regions, with the highest importance score in the EU-15 (\( 0.0690, p < 0.05 \)), indicating a weak but significant impact of price stability on investments. In the OECD (\( 0.0354 \)), G7 (\( 0.0179 \)), and BRICS (\( 0.0064 \)), inflation remains negligible and statistically insignificant (\( p > 0.05 \)), suggesting that price fluctuations are not a critical factor in explaining GFCF under dynamic settings.

\begin{table}
    \centering
    \caption{RandomForest: \textbf{Dynamic Panel Regression} Models. 
    }
    \begin{filecontents*}{RF_lag/RF_result_summary_lag.csv}
Metric,G7,BRICS,EU-15,OECD
LN\_GFCF\_Ratio(t-1),1.2246 ***,0.8789 ***,1.6970 ***,1.7738 ***
,(0.1289),(0.1564),(0.1316),(0.0865)
GDP\_Growth(t-1),0.0280  **,0.0604 ***,0.0461  **,0.0414 
,(0.0037),(0.0161),(0.0040),(0.0028)
LN\_UnEmpl\_Rate(t-1),0.0123 ,0.1351  **,0.0191 ,0.0271 
,(0.0018),(0.0331),(0.0024),(0.0025)
LN\_TAX(t-1),0.1357  **,0.0153 ,0.0956 ***,0.0448 
,(0.0171),(0.0037),(0.0125),(0.0059)
CPI(t-1),0.0179  **,0.0064 ,0.0690  **,0.0354 
,(0.0031),(0.0013),(0.0093),(0.0033)
Observations,123,28,233,376
MSE,0.0002,0.0006,0.0005,0.0006
R-squared,0.9841,0.9758,0.9673,0.9680
Adj. R-squared,0.9835,0.9703,0.9666,0.9676
F-statistic,1355.8744 ***,155.5532 ***,1230.2119 ***,2079.4093 ***
\end{filecontents*}
\csvautotabular[separator=comma]{RF_lag/RF_result_summary_lag.csv}
    \label{tab:rf_dynamic}
    \begin{tablenotes}
        \footnotesize
        \item Note: Value is permutation importance score with standard deviation in brackets below. The p-value significance code of variable importance is presented, and the p-value is based on sequential p-value estimation \cite{hapfelmeier_efficient_2023}. 
        \\
        For brevity, estimates for year dummies and instrument variables are omitted. ***, **, and * denote significance at the 1\%, 5\%, and 10\% levels, respectively. Robust standard errors are shown in parentheses.
        \item Source: Own study.
    \end{tablenotes}
\end{table}

\subsubsection{Key Variable Importance in Random Forest}

In static models, \( GDP\_Growth(t-1) \) demonstrates moderate importance across regions. Its importance is highest in OECD (\( 0.4700, p < 0.01 \)) and BRICS (\( 0.3779, p < 0.01 \)), reflecting its role in driving investments in advanced and emerging markets. In contrast, GDP growth has minimal importance in G7 (\( 0.0753, p > 0.05 \)), indicating its lower influence in mature economies. In dynamic models, the importance of \( GDP\_Growth(t-1) \) declines significantly across all regions. BRICS exhibits the highest score (\( 0.0604, p < 0.05 \)), while G7 (\( 0.0280, p > 0.05 \)), EU-15 (\( 0.0461, p > 0.05 \)), and OECD (\( 0.0414, p > 0.05 \)) show minimal and statistically insignificant effects.

The unemployment rate (\( LN\_UnEmpl\_Rate(t-1) \)) emerges as the dominant factor in BRICS in static settings, with an importance score of \( 0.9080 \) (\( p < 0.01 \)), underscoring labor market vulnerabilities in emerging economies. In OECD (\( 0.5064, p < 0.05 \)), it retains moderate importance. However, in G7 (\( 0.1202, p > 0.05 \)) and EU-15 (\( 0.4213, p > 0.05 \)), unemployment plays a smaller role. In dynamic settings, its influence diminishes, with BRICS showing the highest importance score (\( 0.1351, p < 0.05 \)), while G7 (\( 0.0123, p > 0.05 \)), EU-15 (\( 0.0191, p > 0.05 \)), and OECD (\( 0.0271, p > 0.05 \)) remain statistically insignificant.

Taxation (\( LN\_TAX(t-1) \)) consistently emerges as a critical determinant in advanced economies. In static models, G7 reports the highest importance score (\( 1.5784, p < 0.01 \)), followed by EU-15 (\( 0.4988, p < 0.01 \)) and OECD (\( 0.4437, p < 0.01 \)). In BRICS, taxation shows minimal importance (\( 0.0388, p > 0.1 \)). In dynamic settings, taxation remains significant in G7 (\( 0.1357, p < 0.05 \)) and EU-15 (\( 0.0956, p < 0.05 \)), albeit with reduced importance values. BRICS (\( 0.0153, p > 0.05 \)) and OECD (\( 0.0448, p > 0.05 \)) exhibit weaker and statistically insignificant effects.

Inflation (\( CPI(t-1) \)) demonstrates moderate importance in static models for OECD (\( 0.4002, p < 0.05 \)) and EU-15 (\( 0.3586, p < 0.01 \)), indicating sensitivity to price stability in these regions. In BRICS (\( 0.0281, p > 0.1 \)) and G7 (\( 0.1055, p > 0.05 \)), inflation has minimal importance. In dynamic settings, inflation’s role further diminishes, with the highest score observed in EU-15 (\( 0.0690, p < 0.05 \)). In other regions, such as OECD (\( 0.0354, p > 0.05 \)), G7 (\( 0.0179, p > 0.05 \)), and BRICS (\( 0.0064, p > 0.05 \)), inflation remains negligible.

The lagged \( GFCF \) ratio (\( LN\_GFCF\_Ratio(t-1) \)) is the dominant predictor in dynamic models across all regions, emphasizing the path-dependent nature of investment decisions. Its importance is highest in OECD (\( 1.7738, p < 0.01 \)), followed by EU-15 (\( 1.6970, p < 0.01 \)), G7 (\( 1.2246, p < 0.01 \)), and BRICS (\( 0.8789, p < 0.01 \)).

In static models, \( GDP\_Growth(t-1) \) demonstrates moderate importance across regions. Its importance is highest in OECD (\( 0.4700, p < 0.01 \)) and BRICS (\( 0.3779, p < 0.01 \)), highlighting its role in driving investments in advanced and emerging markets. In G7, GDP growth has minimal importance (\( 0.0753, p > 0.05 \)), reflecting its lower influence in mature economies. EU-15 exhibits a moderate importance score of \( 0.3319 \) (\( p < 0.01 \)). The unemployment rate (\( LN\_UnEmpl\_Rate(t-1) \)) is the dominant factor in BRICS with the highest importance score (\( 0.9080, p < 0.01 \)), underlining the sensitivity of labor markets in emerging economies. In OECD, unemployment retains moderate importance (\( 0.5064, p < 0.05 \)). However, its role in G7 (\( 0.1202, p < 0.05 \)) and EU-15 (\( 0.4213, p > 0.05 \)) is comparatively smaller. Taxation (\( LN\_TAX(t-1) \)) emerges as the most critical factor in advanced economies. G7 shows the highest importance score (\( 1.5784, p < 0.01 \)), followed by EU-15 (\( 0.4988, p < 0.01 \)) and OECD (\( 0.4437, p < 0.01 \)). In BRICS, taxation plays a negligible role (\( 0.0388, p > 0.1 \)). Inflation (\( CPI(t-1) \)) demonstrates moderate importance in OECD (\( 0.4002, p < 0.05 \)) and EU-15 (\( 0.3586, p < 0.01 \)). By contrast, its role is minimal in BRICS (\( 0.0281, p > 0.1 \)) and G7 (\( 0.1055, p > 0.05 \)).
 
In dynamic models, the lagged GFCF ratio (\( LN\_GFCF\_Ratio(t-1) \)) becomes the dominant predictor across all regions. Its importance is highest in OECD (\( 1.7738, p < 0.01 \)), reflecting strong path dependency in advanced economies. The EU-15 follows with a importance score of \( 1.6970 \) (\( p < 0.01 \)), highlighting the role of historical investment trends. In G7, the lagged GFCF ratio shows a importance score of \( 1.2246 \) (\( p < 0.01 \)), while BRICS exhibits a lower but still significant score of \( 0.8789 \) (\( p < 0.01 \)).
The importance of \( GDP\_Growth(t-1) \) diminishes in dynamic settings across all regions. BRICS retains the highest importance score (\( 0.0604, p < 0.05 \)), indicating its continued role in emerging markets. In G7 (\( 0.0280, p > 0.05 \)), EU-15 (\( 0.0461, p > 0.05 \)), and OECD (\( 0.0414, p > 0.05 \)), GDP growth shows minimal and statistically insignificant importance. The unemployment rate (\( LN\_UnEmpl\_Rate(t-1) \)) remains partially relevant in BRICS (\( 0.1351, p < 0.05 \)), while its importance in G7 (\( 0.0123, p > 0.05 \)), EU-15 (\( 0.0191, p > 0.05 \)), and OECD (\( 0.0271, p > 0.05 \)) is negligible and statistically insignificant. Taxation (\( LN\_TAX(t-1) \)) retains moderate importance in G7 (\( 0.1357, p < 0.05 \)) and EU-15 (\( 0.0956, p < 0.05 \)), reflecting the continued role of fiscal policies in advanced economies. In BRICS (\( 0.0153, p > 0.05 \)) and OECD (\( 0.0448, p > 0.05 \)), taxation is less significant. Inflation (\( CPI(t-1) \)) remains the least important variable across regions. The highest importance score is observed in EU-15 (\( 0.0690, p < 0.05 \)), indicating a weak but significant impact of price stability. In OECD (\( 0.0354, p > 0.05 \)), G7 (\( 0.0179, p > 0.05 \)), and BRICS (\( 0.0064, p > 0.05 \)), inflation remains negligible and insignificant.
\subsubsection{Random Forest Evaluation Metrics}\label{RF_Performance_Metrics}

In static settings, the Random Forest models exhibit strong explanatory power across all regions. For the G7, the \( R^2 \) value is \( 0.9841 \), with an Adjusted \( R^2 \) of \( 0.9835 \) and an exceptionally low Mean Squared Error (MSE) of \( 0.0002 \). The \( F \)-statistic is \( 1355.8744 \) (\( p < 0.01 \)), confirming the robustness of the model in explaining investment behavior in advanced economies. In the BRICS, the model achieves an \( R^2 \) of \( 0.9533 \) and an Adjusted \( R^2 \) of \( 0.9452 \), with an MSE of \( 0.0013 \) and an \( F \)-statistic of \( 99.2338 \) (\( p < 0.01 \)), reflecting strong performance despite the small sample size and economic heterogeneity. For the EU-15 and OECD regions, the models demonstrate slightly lower \( R^2 \) values of \( 0.8887 \) and \( 0.8840 \), respectively, with Adjusted \( R^2 \) values of \( 0.8867 \) and \( 0.8827 \). The MSE values for the EU-15 (\( 0.00176 \)) and OECD (\( 0.00218 \)) confirm predictive reliability, with significant \( F \)-statistics of \( 272.3469 \) and \( 410.8423 \) (\( p < 0.01 \)).

In dynamic settings, the inclusion of lagged variables substantially improves model's evaluation metrics. For the G7, the \( R^2 \) remains the highest at \( 0.9841 \), with an Adjusted \( R^2 \) of \( 0.9835 \) and the lowest MSE of \( 0.00018 \). The \( F \)-statistic increases to \( 1355.8744 \) (\( p < 0.01 \)), reinforcing the model's robustness. In the BRICS, the \( R^2 \) improves to \( 0.9758 \), with an Adjusted \( R^2 \) of \( 0.9703 \) and an MSE of \( 0.00065 \). The \( F \)-statistic rises to \( 155.5532 \) (\( p < 0.01 \)), highlighting the importance of lagged investment variables in emerging economies. For the EU-15 and OECD regions, dynamic models show notable improvements with \( R^2 \) values of \( 0.9673 \) and \( 0.9680 \), and Adjusted \( R^2 \) values of \( 0.9666 \) and \( 0.9676 \), respectively. The MSE decreases to \( 0.00052 \) for the EU-15 and \( 0.00060 \) for the OECD. The \( F \)-statistics of \( 1230.2119 \) and \( 2079.4093 \) (\( p < 0.01 \)) confirm the strong explanatory power of lagged predictors in these regions.

The results clearly demonstrate that dynamic models outperform static models across all regions. The G7 group achieves the highest overall performance, reflecting stable and predictable investment dynamics in advanced economies. In contrast, the BRICS group shows the most substantial improvement in dynamic settings, underscoring the role of lagged investment variables in emerging markets. Taxation (\( \text{\textit{LN\_TAX(t}-1)} \)) remains the dominant factor in the G7 and EU-15, emphasizing the critical role of fiscal policies in advanced economies. The Unemployment Rate (\( \text{\textit{LN\_UnEmpl\_Rate(t}-1)} \)) emerges as the primary determinant in BRICS, reflecting the sensitivity of emerging economies to labor market fluctuations. While GDP growth (\( GDP\_Growth(t-1) \)) demonstrates moderate importance in static models, particularly in BRICS and OECD, its influence diminishes in dynamic settings as lagged GFCF ratios (\( \text{\textit{LN\_GFCF\_Ratio(t}-1)} \)) dominate across all regions. Inflation (\( CPI(t-1) \)) shows regional significance in static models for OECD and EU-15 but remains largely insignificant in dynamic settings. 

\subsection{How do non-linear Random Forest models improve the understanding of relationships between macroeconomic predictors (e.g., GDP growth, taxation, and unemployment) and GFCF compared to linear static and dynamic regression models?}

This section compares the performance of linear regression models (static and dynamic) with non-linear Random Forest models in both static and dynamic settings. The comparison focuses on \( R^2 \), Adjusted \( R^2 \), Mean Squared Error (MSE), and \textit{F}-statistics as indicators of model fit and robustness.

\subsubsection{Key Findings Across Country Groups}

For the G7, both static and dynamic models reveal notable patterns. In Static Panel Regression, \textit{GDP\_ Growth} (\( 0.0112, p < 0.01 \)) and \textit{Unempl\_Rate} (\( -0.1354, p < 0.01 \)) emerge as significant predictors, while Taxation has the largest negative effect (\( -0.4551, p < 0.01 \)). However, Random Forest (Static) highlights Taxation as the most important variable (\( 1.5784, p < 0.01 \)), followed by the \textit{Unempl\_Rate} (\( 0.1202 \)). In Dynamic Panel Regression, the inclusion of lagged GFCF (\( 0.9595, p < 0.01 \)) dominates the model, reducing the significance of other predictors. The Random Forest (Dynamic) model confirms this, assigning the highest importance score to lagged GFCF (\( 1.2246, p < 0.01 \)), with \textit{GDP\_Growth} and other variables playing minimal roles. Random Forest models consistently outperform panel regressions, achieving higher \( R^2 \) (\( 0.9841 \)) and lower MSE (\( 0.00018 \)).

In the BRICS group, economic heterogeneity leads to different findings. Static Panel Regression shows that \textit{GDP\_Growth} (\( 0.0128, p < 0.01 \)) and \textit{Unempl\_Rate} (\( -0.2250, p < 0.01 \)) are key predictors, while Taxation (\( 0.4077, p < 0.1 \)) has a surprising positive impact. In contrast, Random Forest (Static) identifies the \textit{Unempl\_Rate} as the dominant predictor (\( 0.9080, p < 0.01 \)), followed by \textit{GDP\_Growth} (\( 0.3779, p < 0.01 \)). In Dynamic Panel Regression, lagged GFCF (\( 0.9238, p < 0.01 \)) becomes the most significant factor, reflecting investment inertia. Random Forest (Dynamic) confirms this with an importance score of \( 0.8789 \), while GDP\_Growth remains relevant (\( 0.0604, p < 0.05 \)). Despite the small sample size, Random Forest achieves higher \( R^2 \) (\( 0.9758 \)) compared to panel regressions (\( 0.5720 \)).

For the EU-15, Static Panel Regression identifies moderate significance of \textit{GDP\_Growth} (\( 0.0088, p < 0.01 \)) and \textit{Unempl\_Rate} (\( -0.1280, p < 0.01 \)), while Taxation (\( -0.2602, p < 0.05 \)) also plays a negative role. Random Forest (Static) emphasizes Taxation (\( 0.4988, p < 0.01 \)) and the \textit{Unempl\_Rate} (\( 0.4213 \)) as key predictors. In Dynamic Panel Regression, lagged GFCF (\( 0.9594, p < 0.01 \)) dominates, with \textit{GDP\_Growth} (\( 0.0068, p < 0.01 \)) retaining weak significance. Random Forest (Dynamic) achieves stronger performance, with an \( R^2 \) of \( 0.9673 \) and an importance score of \( 1.6970 \) for lagged GFCF. Compared to panel regressions (\( R^2 = 0.2733 \)), Random Forest significantly enhances model fit.

In the OECD group, Static Panel Regression yields limited explanatory power, with \textit{GDP\_Growth} (\( 0.0093, p < 0.01 \)) and Inflation (\( 0.0123, p < 0.01 \)) showing minor significance. Taxation (\( -0.2330, p < 0.01 \)) has a notable negative effect. However, Random Forest (Static) highlights Taxation (\( 0.4437, p < 0.01 \)) and \textit{Unempl\_Rate} (\( 0.5064 \)) as critical variables, achieving a much higher \( R^2 \) (\( 0.8840 \)). In Dynamic Panel Regression, lagged GFCF dominates (\( 0.9856, p < 0.01 \)), reducing the relevance of other predictors. Random Forest (Dynamic) confirms this, assigning the highest importance to lagged GFCF (\( 1.7738, p < 0.01 \)). The model achieves an \( R^2 \) of \( 0.9680 \), far exceeding the static panel regression's \( 0.0992 \).

Random Forest models reveal non-linear interactions and hidden patterns, particularly for GDP growth, which are not adequately captured by static and dynamic panel regressions. The reduced importance of GDP growth in dynamic Random Forest models suggests that its relationship with GFCF is mediated by lagged investment values in a more complex, non-linear manner. This relationship is less evident in linear models. Taxation consistently emerges as a critical determinant of GFCF across both static and dynamic settings, particularly in the G7, EU-15, and OECD regions. Static panel regressions reveal its negative impact on GFCF, while Random Forest assigns it a high level of importance, underscoring its role as a fiscal constraint or enabler in advanced economies. The unemployment rate plays a prominent role in BRICS and EU-15. In static settings, it is the dominant predictor for BRICS, reflecting the sensitivity of emerging markets to labor market fluctuations. Random Forest models better capture this relationship compared to linear models, highlighting the added value of non-linear approaches. Inflation, on the other hand, exhibits limited importance across all regions. It shows minor significance in the OECD within static models, but its relevance further diminishes in dynamic settings, suggesting that price stability exerts minimal influence on capital formation relative to other predictors. Finally, lagged GFCF emerges as the most influential predictor in dynamic models across all regions, emphasizing the persistence and path dependency of investments. Random Forest models confirm its dominant role, particularly in the G7, EU-15, and OECD, where historical investment levels significantly shape current capital formation dynamics.

\subsubsection{Comparison of Model Performance: Linear vs. Non-linear Models}

In static settings, Random Forest models outperform static panel regression models across all regions. The results demonstrate a significantly higher explanatory power and lower error values in the Random Forest models.

In the G7, the static panel regression achieves an \( R^2 \) of \( 0.4585 \), with an Adjusted \( R^2 \) of \( 0.4402 \) and an \textit{F}-statistic of \( 23.7084 \) (\( p < 0.01 \)). In contrast, the Random Forest model achieves near-perfect explanatory power, with an \( R^2 \) of \( 0.9841 \), an Adjusted \( R^2 \) of \( 0.9835 \), and an exceptionally low Mean Squared Error (MSE) of \( 0.0002 \). This highlights the Random Forest model's ability to capture subtle variations and complex interactions in advanced and stable economies like the G7.

For the BRICS group, the static panel regression achieves the highest \( R^2 \) among linear models at \( 0.5720 \), with an Adjusted \( R^2 \) of \( 0.5061 \) and an F-statistic of \( 7.6840 \) (\( p < 0.01 \)). However, the Random Forest model far outperforms it, achieving an \( R^2 \) of \( 0.9533 \), an Adjusted \( R^2 \) of \( 0.9452 \), and an MSE of \( 0.0013 \). This substantial improvement underscores the importance of non-linear relationships in emerging economies, where heterogeneity and dynamic growth patterns make linear models less effective.

In the EU-15, the static panel regression explains only \( 27.33\% \) of the variation in \( \text{\textit{GFCF\_Ratio}} \) (\( R^2 = 0.2733 \)), with an Adjusted \( R^2 \) of \( 0.2608 \) and an F-statistic of \( 20.5922 \) (\( p < 0.01 \)). The Random Forest model demonstrates a significant improvement, achieving an \( R^2 \) of \( 0.8887 \), an Adjusted \( R^2 \) of \( 0.8867 \), and an MSE of \( 0.00176 \). These results suggest that linear models struggle to capture the economic dynamics in this economically integrated yet heterogeneous group, while Random Forest better identifies complex patterns.

In the OECD, despite the largest sample size (\( N = 388 \)), the static panel regression yields the lowest \( R^2 \) (\( 0.0992 \)) and Adjusted \( R^2 \) (\( 0.0898 \)), with an F-statistic of \( 9.8855 \) (\( p < 0.01 \)). In contrast, the Random Forest model achieves a far superior performance, with an \( R^2 \) of \( 0.8840 \), an Adjusted \( R^2 \) of \( 0.8827 \), and an MSE of \( 0.00218 \). The large improvement indicates that Random Forest can effectively handle the economic diversity and structural complexities of OECD countries, where linear models fail to capture these relationships.

The comparison clearly demonstrates the superiority of Random Forest models over static linear regression models across all regions. This difference is particularly notable in OECD and BRICS, where the Random Forest achieves significantly higher \( R^2 \) values despite the OECD's large sample size and the BRICS' economic heterogeneity. These findings highlight the capacity of non-linear models to uncover complex relationships that static linear models cannot adequately capture.

\subsubsection{Comparison of Model Performance: Dynamic Regression Models and Random Forest in Dynamic Settings}

In dynamic settings, the inclusion of lagged \( \text{\textit{GFCF\_Ratio}} \) improves model performance in both linear and non-linear approaches. However, Random Forest models still achieve higher \( R^2 \), lower MSE, and stronger F-statistics across all regions.

For G7 group, in the dynamic panel regression, lagged \( \text{\textit{GFCF\_Ratio}} \) is highly significant (\( 0.9595, p < 0.01 \)), confirming the persistence of investment patterns in mature economies. However, the Random Forest model achieves superior performance, with an importance score of \( 1.2246 \) for lagged \( \text{\textit{GFCF\_Ratio}} \), an \( R^2 \) of \( 0.9841 \), and an exceptionally low MSE of \( 0.00018 \). This highlights the ability of Random Forest to better capture complex relationships within stable economies.

In the BRICS, dynamic panel regression results indicate that lagged \( \text{\textit{GFCF\_Ratio}} \) remains significant (\( 0.9238, p < 0.01 \)), underlining the investment inertia in emerging markets. However, the RF model demonstrates a clear advantage, achieving \( R^2 = 0.9758 \), an Adjusted \( R^2 \) of \( 0.9703 \), and a lower MSE of \( 0.00065 \). The importance score of \( 0.8789 \) for lagged \( \text{\textit{GFCF\_Ratio}} \) further confirms its dominance. The greater explanatory power of RF highlights its ability to account for the economic heterogeneity of BRICS countries.

In the EU-15 dataset, dynamic linear regression shows strong performance with lagged \( \text{\textit{GFCF\_Ratio}} \) (\( 0.9594, p < 0.01 \)). Nevertheless, the RF improves results, achieving \( R^2 = 0.9673 \), Adjusted \( R^2 = 0.9666 \), and an MSE of \( 0.00052 \). The importance score of \( 1.6970 \) for lagged \( \text{\textit{GFCF\_Ratio}} \) underscores its role as the primary predictor. The clear improvement in performance suggests that the RF is better equipped to capture subtle variations in investment behavior within the region.

In the OECD, in the dynamic panel regression, lagged \( \text{\textit{GFCF\_Ratio}} \) dominates with a coefficient of \( 0.9856 \) (\( p < 0.01 \)), reflecting strong investment inertia in diverse and advanced economies. The RF model, however, achieves slightly higher explanatory power, with \( R^2 = 0.9680 \), an Adjusted \( R^2 \) of \( 0.9676 \), and an MSE of \( 0.00060 \). The importance score of \( 1.7738 \) for lagged \( \text{\textit{GFCF\_Ratio}} \) highlights the model's precision in capturing complex patterns within heterogeneous OECD countries.

The comparison of model performance highlights the superior accuracy and robustness of Random Forest models over linear regression models in both static and dynamic settings. While dynamic linear models improve significantly over their static counterparts, Random Forest models deliver the highest \( R^2 \), lower MSE, and strong F-statistics across all regions. These differences are particularly pronounced in BRICS and OECD regions, where the non-linear Random Forest approach demonstrates its ability to capture intricate relationships and address economic heterogeneity. The results suggest that Random Forest models provide a more accurate representation of investment dynamics, particularly when path-dependent variables like lagged GFCF play a dominant role.

\section{Discussion}

\subsection{Regional Variability in the Influence of GDP Growth on GFCF}

The results confirm the [H1]. The study's findings reveal a robust relationship between GDP growth and GFCF across both developed (G7, EU-15, OECD) and emerging (BRICS) markets. Specifically, static and dynamic regression models consistently indicate that GDP growth exerts a statistically significant and positive impact on GFCF, suggesting that economic expansion serves as a reliable driver of corporate investment. These results align with other studies, which emphasize the role of GDP as a primary determinant of capital formation \cite{jorgenson_capital_1963}, and recent findings (e.g., \cite{le_foreign_2024}, \cite{pasara_causality_2020}, which affirm the bidirectional relationship between economic growth and capital formation. Notably, while developed markets show a stronger response to GDP growth in terms of corporate investment, emerging markets like BRICS, though positively influenced, experience a comparatively moderated effect. This disparity may stem from structural and economic limitations that characterize emerging economies, as suggested by \cite{feldstein_domestic_1980}, who highlight the national savings constraints that can impact capital formation in developing countries.

The finding that developed markets exhibit a stronger GDP-GFCF relationship is supported by studies on regional capital formation differences. For example, \cite{blanchard_lectures_1989} and \cite{de_long_equipment_1990} highlight that developed economies typically enjoy more stable investment environments and easier access to financing, which may amplify the effects of GDP growth on corporate investment. Conversely, \cite{xiang_time-varying_2021} show that while BRICS countries experience positive impacts from GDP growth, these effects may be tempered by structural constraints, including financial market limitations and economic volatility, common in emerging economies. Furthermore, \cite{apergis_renewable_2010} and \cite{world_bank_world_2018} illustrate how sector-specific responses to economic growth in emerging regions can yield varied GFCF outcomes, suggesting that the GDP-GFCF link in BRICS might be concentrated in sectors with greater capital sensitivity.

The dynamic panel model findings further underscore the path-dependent nature of GFCF, where previous investment levels significantly influence current GFCF, especially in developed economies. This trend implies that firms in developed regions benefit from a stable investment climate and a supportive financial infrastructure that facilitates reinvestment. The results echo \cite{borensztein_how_1998} and \cite{blomstrom_multinational_1998}, who demonstrate that in economies with strong financial structures, past investments catalyze future GFCF by reinforcing corporate confidence and stability. This path dependency is observed less strongly in emerging markets, where economic conditions are less predictable, and firms may face greater reinvestment challenges due to financial constraints.

Overall, these results corroborate [H1] by demonstrating that GDP growth indeed fosters corporate investment across both developed and emerging markets. However, for emerging markets to fully capitalize on economic growth, complementary policies that address investment barriers may be essential. For instance, addressing structural financial limitations, as suggested by \cite{feldstein_domestic_1980}, could enhance the efficiency of capital formation in these regions. Thus, while GDP growth is a critical factor in driving corporate investment, the strength and consistency of this relationship vary by economic maturity and market stability, as highlighted in the literature by \cite{de_long_equipment_1990} and \cite{xiang_time-varying_2021}.

In developed regions (G7, EU-15, OECD), policymakers should strengthen the GDP-GFCF link by offering tax incentives to encourage reinvestment and improving access to finance for long-term projects. Public infrastructure investments can complement private capital formation, creating a feedback loop to support growth. Additionally, focusing on high-potential sectors like renewable energy and technology can maximize GFCF’s impact on sustainable development. Emerging markets (BRICS) need policies to address financial inefficiencies and low savings rates. Expanding microfinance, creating sovereign wealth funds, and enhancing FDI frameworks can help bridge investment gaps. Strengthening institutions to reduce volatility and fostering investor confidence are key. Sector-specific incentives for manufacturing and green industries can further boost GFCF’s response to economic growth. Across all regions, coordinated policies for trade blocs can harmonize investment rules, enhancing cross-border capital flows. Governments should also recognize GFCF’s path dependency by offering reinvestment incentives, ensuring stability and sustained growth.

\subsection{The Role of Other Macroeconomic Predictors than GDP Growth in Emerging vs. Developed Economies}

The analysis corroborates hypothesis [H2]. The findings reveal that the relationship between GDP growth and GFCF is more robust in emerging markets (BRICS) compared to developed economies (G7, OECD, EU-15). These results align with \cite{jorgenson_capital_1963}, who emphasized GDP growth as a key driver of investment, and \cite{xiang_time-varying_2021}, who highlighted the stronger impacts of GDP growth on investment in BRICS compared to developed markets. Moreover, \cite{barro_economic_1989} underscores the positive correlation between GDP growth and investment across diverse economic contexts, further substantiating these findings. In developed economies, the moderated GDP-GFCF correlation observed here reflects findings by \cite{blanchard_lectures_1989} and \cite{de_long_equipment_1990}, who identified greater stability in investment dynamics due to well-developed financial infrastructures. This is consistent with findings from \cite{boamah_financial_2018}, which emphasize that GFCF significantly contributes to economic growth in Asian economies, underscoring the importance of capital formation in fostering growth in emerging markets. However, \cite{devereux_taxes_1998} note that the efficiency of this relationship in developed markets is contingent on fiscal and policy frameworks, particularly corporate tax incentives.
However, the study also highlights the role of financial depth in potentially impeding growth, which could moderate the GDP-GFCF relationship.
Additionally, the existence of Global Financial Cycles (GFCy), as discussed by\cite{dua_dynamics_2024}, constrains policymakers’ ability to shield domestic economies from global financial trends, indicating that emerging markets are particularly sensitive to external shocks.

Taxation emerged as a critical determinant with contrasting effects across regions. In advanced economies (G7, EU-15), higher taxes significantly deter GFCF, reflecting fiscal constraints, consistent with \cite{hall_tax_1967}, who demonstrated the adverse impact of tax burdens on corporate investment, and \cite{elif_arbatli_c_economic_2011}, who highlights that reducing corporate tax rates is a vital strategy for attracting foreign direct investment (FDI) and enhancing domestic investment. Conversely, in BRICS, taxation occasionally supports GFCF, potentially through targeted fiscal incentives, echoing findings by \cite{beyer_early_2021} and \cite{crawford_effect_2024}, who showed how well-designed tax incentives could stimulate investment. 

Unemployment was found to influence GFCF negatively in both developed and emerging economies, albeit more significantly in the former. In emerging markets, labor market dynamics have a direct effect on investment confidence, emphasizing the importance of an active labor force in sustaining economic growth, as highlighted by \cite{boamah_financial_2018}. This finding aligns with \cite{blanchard_regional_1992}, who discussed the relationship between unemployment and reduced investment due to lower demand.  Additionally, \cite{phelps_money-wage_1968} noted that capital-intensive investments might create short-term employment challenges, which could partly explain the varying impacts across regions. Contrarily, \cite{pasara_causality_2020} suggest a potential counter-effect in some contexts, where firms may increase capital formation during periods of high unemployment to capitalize on lower labor costs.

Beyond taxation and unemployment, additional variables such as inflation and some of the experimental variables (control or instrumental), e.g. economic policy uncertainty, were also observed to influence GFCF variably. Developed markets showed higher sensitivity to inflation stability, as highlighted by \cite{montes_effects_2021}, who emphasized the importance of macroeconomic predictability in maintaining business confidence. Meanwhile, emerging markets exhibited more robust GFCF responses to GDP growth, with relatively muted impacts from inflation and uncertainty, consistent with the observations of \cite{feldstein_domestic_1980} regarding structural savings constraints. Additionally, \cite{boamah_financial_2018} document that financial depth, while essential for investment, may hinder growth in the absence of robust regulatory frameworks, particularly in emerging markets. \cite{rey_dilemma_2015} underscores the critical role of GFCy in shaping investment trends, especially in economies exposed to external shocks. Moreover, \cite{dua_dynamics_2024} underline the critical impact of global risk factors and the GFCy on capital flow volatility, showing that domestic economies' ability to manage these influences significantly determines GFCF outcomes.

For developed markets, policies should focus on tax reform, financial stability, and infrastructure investment to sustain the robust GDP-GFCF relationship. Promoting macroeconomic predictability will also enhance business confidence and long-term investment. In emerging markets, addressing structural constraints like weak financial frameworks and fostering investor confidence through institutional reform are critical. Sector-specific incentives targeting manufacturing and technology can further enhance GFCF responsiveness to GDP growth. Across all regions, coordinated international strategies to manage GFCy can improve resilience and optimize GFCF allocation.

\subsection{Non-linear Relationships and the Role of Random Forest Models}

Random Forest confirms [H3] by identifying non-linear interactions and capturing complex relationships between variables. As emphasized by \cite{breiman_random_2001}, ensemble learning methods like RF provide robust predictions by aggregating outputs from multiple decision trees. This characteristic allows these models to handle economic datasets with high dimensionality and heterogeneity, making them particularly suitable for analyzing macroeconomic predictors in both developed and emerging economies (e.g., OECD and BRICS). For instance, RF identifies lagged GFCF as a dominant predictor in dynamic settings, highlighting the persistence of investment behaviors over time. This finding aligns with earlier studies emphasizing the importance of path dependence in investment decisions \citep{blundell_initial_1998,borensztein_how_1998}.

Comparing evaluation metrics reveals the superior accuracy of RF over traditional linear regressions. For example, in OECD countries, RF achieved an \( R^2 \) of 0.968, far exceeding the 0.099 reported for static panel regressions. Similarly, in BRICS, where economic heterogeneity poses challenges for linear models, RF recorded an \( R^2 \) of 0.976, underscoring its capability to capture complex growth patterns. These results echo findings by \cite{goulet_coulombe_macroeconomy_2024}, who demonstrated the efficacy of RF in forecasting macroeconomic indicators, particularly in heterogeneous datasets.

The study uniquely employs RF not for prediction but to model non-linear relationships and compare its performance with traditional linear regressions. This approach highlights RF’s versatility in uncovering hidden dynamics, such as the reduced direct importance of GDP growth in dynamic models. Instead, lagged GFCF consistently emerged as the most significant factor across regions, underscoring the inertia in investment patterns. This supports \cite{blanchard_regional_1992}, who noted that prior investments often dictate future capital formation. Taxation retained prominence in advanced economies, while unemployment emerged as critical in emerging markets, consistent with findings by \cite{beyer_early_2021} and \cite{pasara_causality_2020}.

Despite its strengths, RF highlighted challenges, such as the limited role of inflation in influencing GFCF, particularly in dynamic settings. While \cite{montes_effects_2021} identified policy uncertainty and inflation as key factors affecting investment confidence, their relative importance diminished in RF analyses, especially in regions like BRICS. This divergence suggests that while inflation and uncertainty affect short-term decisions, their long-term influence may be mediated by other factors.

The ability of RF to capture time-lagged effects further enhances its utility. By incorporating temporal dependencies, these models reveal that investment decisions are heavily influenced by past economic conditions. This insight is critical for policymakers aiming to design effective interventions. Advanced economies could focus on maintaining stable tax policies and incentivizing reinvestment, while emerging markets might prioritize addressing unemployment to stimulate capital formation. Ultimately, this study demonstrates RF’s value as a diagnostic tool for uncovering complex macroeconomic dynamics, extending its traditional role in prediction to modeling and comparative analysis.

The implementation of the p-value importance algorithm for Random Forest was parallelized in Python to improve computational efficiency while maintaining statistical validity. The parallel processing approach distributes the permutation testing across multiple CPU cores, significantly reducing computation time for large datasets. This implementation follows the theoretical framework for testing variable importance measures (VIMP) in Random Forests, incorporating sequential testing methods like SPRT (Sequential Probability Ratio Test) and SAPT (Sequential Approximation to Permutation Test). The algorithm maintains the statistical rigor of the original methodology while leveraging Python's multiprocessing capabilities to handle the computationally intensive permutation procedures, particularly beneficial when analyzing extensive macroeconomic datasets across multiple regions.

\section{Limitations}\label{limitations}

\paragraph{Data Limitations.}
The limited number of observations in certain datasets, particularly for BRICS (28 observations), raises concerns about the generalizability of the results. The small sample size, combined with a high \( R^2 \) value (0.953 for Random Forest and 0.572 for static panel regression), may indicate potential overfitting, where the model may fit the sample data too closely at the expense of broader applicability. In contrast, datasets with larger sample sizes, such as OECD (376 observations), yield more reliable estimates. However, the slightly lower \( R^2 \) values in these datasets reflect the increased heterogeneity among member countries, underscoring the challenge of modeling diverse economic structures within a single framework.

\paragraph{Scope of Variables.}
This study predominantly emphasizes macroeconomic indicators such as GDP growth, taxation, unemployment, and inflation. However, other critical factors, including sector-specific and region-specific investments, were not thoroughly examined, thereby limiting the scope of the analysis. Additionally, while elements such as EPU and GFCy were acknowledged as important influences, they were not explicitly incorporated into the models. This omission restricts the models' capacity to comprehensively capture the effects of external shocks and global interdependencies on economic outcomes.

\paragraph{Methodological Limitations.}
While Random Forest model outperform linear regressions in capturing non-linear relationships, they present several important limitations. They lack the straightforward interpretability of traditional econometric models, and since they are based on machine learning rather than classical statistical theory, conventional inference metrics like p-values must be computed through simulation methods. Furthermore, while we adapted the RF framework for panel data analysis, this adaptation represents a departure from the model's original design for cross-sectional prediction tasks. The black-box nature of RF models internal structure may limit their applicability in policy-making contexts, where transparency is required. Additionally, RF's reliance on permutation importance for variable selection, while useful for identifying predictive relationships, does not establish causality in the same way that carefully specified econometric models can through their coefficients and statistical significance metrics.


\paragraph{Dynamic Panel Analysis Challenges.}
In dynamic settings, the dominance of lagged GFCF highlights investment inertia but may downplay the nuanced roles of other macroeconomic predictors in the short term. Additionally, limited exploration of interaction effects between variables in the dynamic RF model might overlook how combined predictors influence GFCF.

\section{Future Directions}\label{future_directions}

\paragraph{Expanding Datasets.} Future research could enhance the accuracy and robustness of linear models, particularly for underrepresented regions such as the BRICS. Incorporating sector-specific data or disaggregated components of GFCF could yield more detailed insights into investment patterns and regional dynamics. Furthermore, increasing the number of observations in smaller datasets by transitioning from annual to higher-frequency data, such as quarterly observations, may address issues related to sample size limitations. Alternatively, leveraging advanced techniques such as machine learning, as demonstrated in this study, offers a viable solution. Approaches like data augmentation can effectively increase the number of observations, mitigating the challenges posed by small sample sizes while preserving the reliability of the models. 

\paragraph{Replacing a Set of Variables.} An equally intriguing direction for future research involves exploring an alternative, less conventional set of regressors that have thus far been treated as experimental variables. In this study, we employed control and instrumental variables such as Foreign Direct Investment (FDI), the Human Development Index (HDI), the Economic Policy Uncertainty (EPU) Index, and the Gini Index. However, juxtaposing these variables with more granular data, such as sector-specific indicators or the type and scope of corporate investments, could offer deeper insights into the determinants of GFCF. Such an approach may enhance the explanatory power of models and reveal previously unrecognized relationships.

\paragraph{Model Enhancements.}
Further research could explore hybrid approaches that combine Random Forests with interpretable machine learning techniques, particularly SHAP (SHapley Additive exPlanations) values and LIME (Local Interpretable Model-agnostic Explanations), to enhance model transparency for policy applications (see \cite{panda_explainable_2023}. Additionally, developing region-specific RF models could better capture unique economic structures and institutional frameworks across different countries or regions.

\paragraph{Temporal and Interaction Effects.}
The analysis could be extended to examine complex interaction effects between key economic variables, such as how GDP growth moderates the relationship between taxation policies and GFCF, or how unemployment rates influence the effectiveness of fiscal policies. To better capture temporal dynamics, future work could incorporate time series components through methods such as temporal fusion transformers \citep{laborda_multi-country_2023} or RF-based time series models \citep{lin_random_2017,  shiraishi_time_2024}, which might provide more accurate predictions while maintaining the advantages of tree-based approaches.

\section{Concluding comments}\label{Conclusions}
By bridging traditional econometric methods with advanced machine learning, this study provides a nuanced understanding of the factors driving GFCF, across developed and emerging markets.
The findings confirm that GDP growth positively influences GFCF across all regions, with a more pronounced impact in developed economies (G7, EU-15, OECD). This is attributed to stable investment climates and robust financial systems. In emerging markets (BRICS), however, constraints such as lower national savings and economic heterogeneity weaken the GDP-GFCF relationship. 
Beyond GDP growth, factors like taxation and unemployment also play crucial roles. Developed markets show greater sensitivity to taxation, whereas unemployment and labor dynamics significantly influence GFCF in emerging economies.
GDP growth has reduced importance in non-linear models compared to linear regressions, showing that traditional models may overstate its direct impact.
Other macroeconomic factors exhibit non-linear effects, which are better captured by Random Forest than by panel regressions. Taxation plays a more significant role in G7 and EU-15 economies but has a negligible effect in BRICS. Unemployment Rate is crucial in BRICS but less relevant in developed economies, highlighting regional heterogeneity in the labor market's effect on investment.

Random Forest clearly outperforms traditional models by identifying how multiple macroeconomic factors interact in ways that linear models cannot capture. GDP growth's effect on investment is not constant but varies depending on other macroeconomic conditions. The dominant role of lagged GFCF suggests that investment behavior is better explained by past investments rather than GDP growth alone.

Our work advances the application of Random Forests in macroeconomic analysis by developing a parallelized p-value importance algorithm, enhancing efficiency for large datasets while ensuring statistical rigor through sequential testing methods like SPRT and SAPT. This approach distinguishes our study from traditional applications by uncovering non-linear interactions and ranking predictor importance in high-dimensional macroeconomic data. The findings emphasize the need for region-specific policy design and the integration of advanced methodologies for more robust economic analysis.

\appendix
\appendix

\printcredits

\bibliographystyle{cas-model2-names}

\bibliography{zotero}

\newpage

\bio{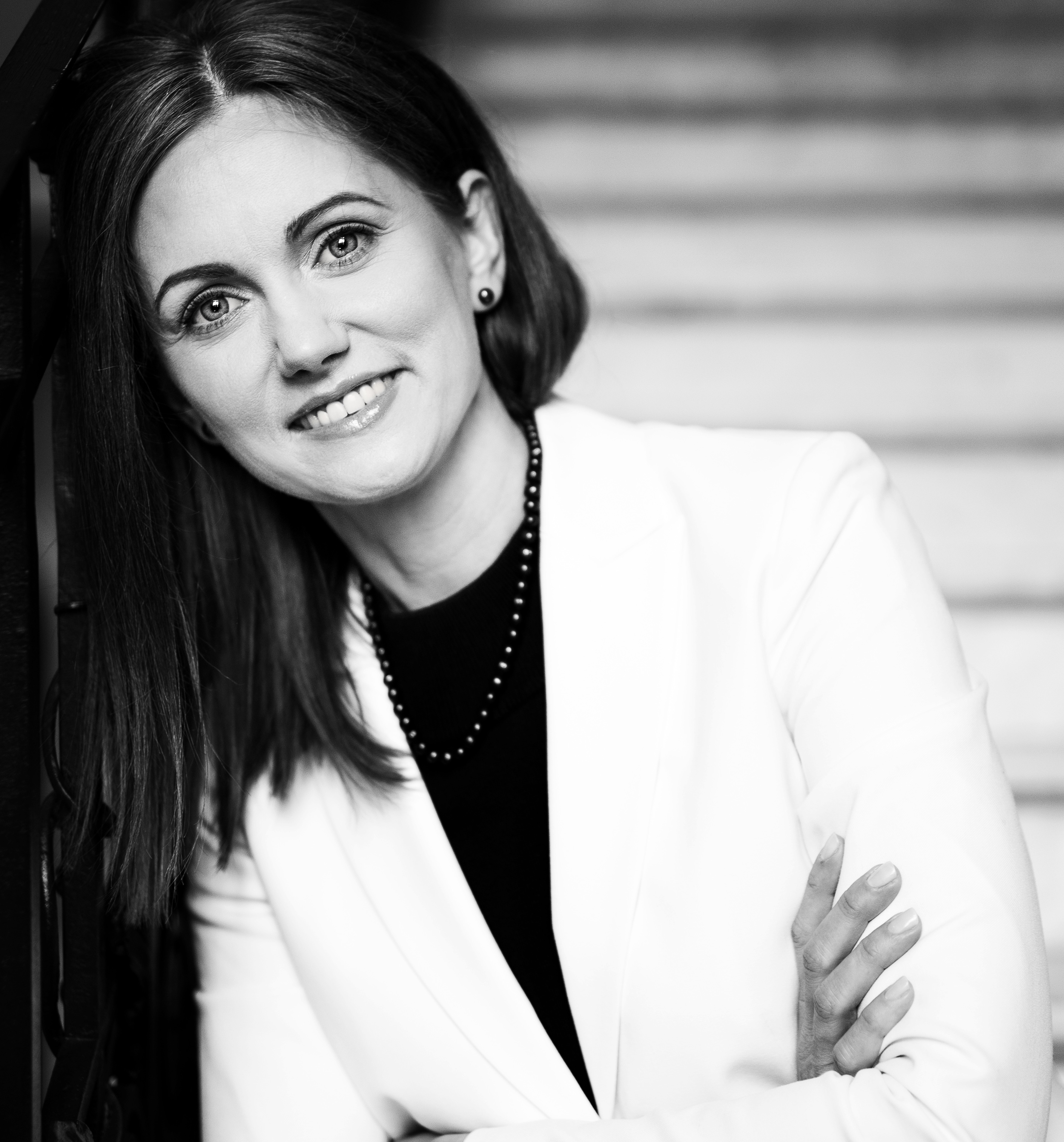}
\textbf{Alina Landowska, PhD}\\ 
I am an Associate Professor at SWPS University and a Research Fellow at the Centre for the World Economy at Stefan Cardinal Wyszyński University in Warsaw, Poland. I serve as a Polish expert at the International Council for Small Business and previously represented Employers of Poland at the Business and Industry Advisory Committee (BIAC) of the OECD. I have been a scholarship recipient at institutions such as the Tantur Ecumenical Institute in Jerusalem (Israel/USA), the Baltic University Programme (Sweden), and Roskilde University (Denmark). My research interests focus on narrative economics, international economics, and computational linguistics, with particular emphasis on text data mining.
\endbio

\bio{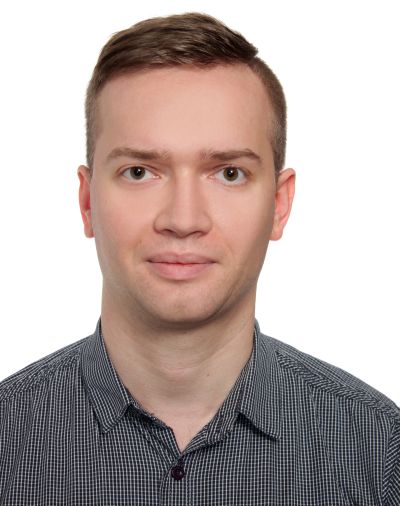}
\textbf{Robert A. Kłopotek, PhD}\\ 
I am an Associate Professor and Deputy Director of Institute of Computer Science at Cardinal Stefan Wyszyński University in Warsaw, Poland. I also work in an international company as a senior data scientist. I graduated PhD studies at Institute of Computer Science, Polish Academy of Sciences. As a data analyst, I utilize scientific knowledge related to statistical data analysis and machine learning algorithms, which I combine with business practice. My professional interests include social network analysis, LLMs, data mining, data analysis, application of GPGPU in selected machine learning algorithms, and theoretical basis for machine learning algorithms.
\endbio

\bio{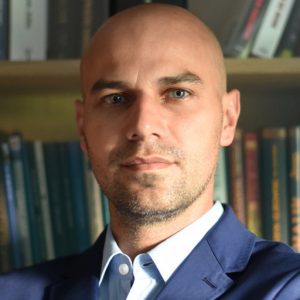}
\textbf{Dariusz Filip, PhD}\\
I am a Doctor of Economic Sciences with a specialization in finance. My area of academic interest focuses on evaluating the efficiency of asset management in financial institutions. I am the author of numerous publications in both Polish and international journals, including "Finance a úvěr-Czech Journal of Economics and Finance", "Baltic Journal of Economics", "Prague Economic Papers", and "Financial Assets and Investing".
\endbio

\bio{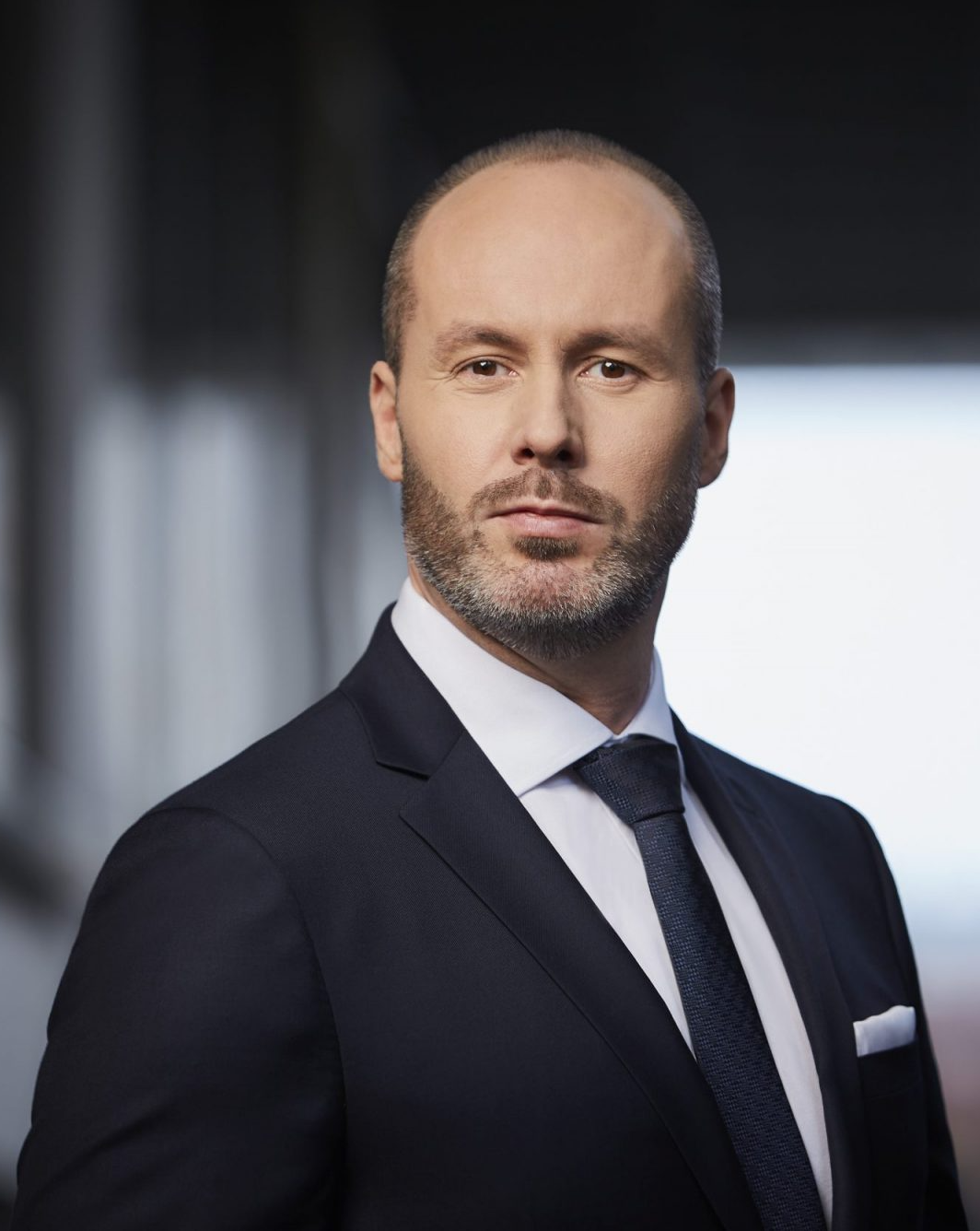}
\textbf{Prof. Konrad Raczkowski, PhD}\\ 
I am a Professor of Economic Sciences with a post-doctoral degree in public institution management and Director of the World Economy Center at Cardinal Stefan Wyszyński University. My specializations include economic system management, public finance, informal economy, and corporate restructuring, having served in leadership positions at academic institutions and as a member of the National Development Council under the President of Poland. I have held executive and advisory roles in companies listed on the Warsaw, New York, and London Stock Exchanges, and served as Deputy Minister of Finance while collaborating with international organizations such as the European Anti-Fraud Office. Throughout my career, I have authored over 120 scientific publications and received numerous honors, including gold medals from the Minister of Justice and Minister of Finance, as well as the Polish Academy of Sciences President's award for outstanding achievements in finance.
\endbio

\end{document}